\documentclass[12]{article}

\usepackage{graphicx}
\usepackage{amsfonts}
\usepackage{amssymb}
\usepackage{wasysym}
\usepackage{makeidx}
\usepackage{multicol}

\newtheorem{thm}{Theorem}
\newtheorem{lem}{Lemma}
\newtheorem{tab}{Table}
\newtheorem{fig}{Figure}
\newtheorem{prop}{Proposition}

\def\leurre{\noindent\leftskip0pt\small\baselineskip 10pt}

\def\encadre#1#2{%
\setbox100=\hbox{\kern#1{#2}\kern#1}
\dimen100=\ht100 \advance \dimen100 by #1
\dimen101=\dp100 \advance \dimen101 by #1
\setbox100=\hbox{\vrule height \dimen100 depth \dimen101\box100\vrule}
\setbox100=\vbox{\hrule\box100\hrule}
\advance \dimen100 by .4pt \ht100=\dimen100
\advance \dimen101 by .4pt \dp100=\dimen101
\box100
\relax
}

\def\ligne#1{\hbox to \hsize{#1}}
\def\PlacerEn#1 #2 #3 {\rlap{\kern#1\raise#2\hbox{#3}}}

\setbox221=\hbox{\includegraphics{cercle_L20.eps}}

\def\encercle#1#2{\hbox{\raise-5pt\copy221\hskip#2#1}}

\title{A decidability result for the halting of cellular automata in the pentagrid
}
\author{Maurice \sc{Margenstern},\\ 
LGIMP, D\'epartement Informatique et Applications,\\
Universit\'e de Lorraine\\
3 rue Augustin Fresnel, BP 45112\\
57073 Metz C\'edex 03, France,\\
{\tt margenstern@gmail.com}
}
\begin{document}
\maketitle

\vskip 10pt
{\bf Abstract}{\small\\
In this paper, we investigate the halting problem for deterministic cellular automaton in 
the pentagrid.
We prove that the problem is decidable when the cellular automaton starts its computation 
from a
finite configuration and when it has at two states, one of them being a quiescent
state.
}

{\bf Keywords}: tilings, hyperbolic geometry, cellular automata, halting problem,
decidability 
\vskip 10pt

\def\cqfd{\hbox{\kern 2pt\vrule height 6pt depth 2pt width 8pt\kern 1pt}}
\def\Hii{$I\!\!H^2$}
\def\Hiii{$I\!\!H^3$}
\def\Hiv{$I\!\!H^4$}
\def\norm{\hbox{$\vert\vert$}}
\section{Introduction}\label{intro}

   Let us start with generalities about cellular automata. A cellular automaton is
defined by two basic objects: the space of its cells and the finite automaton, a copy of 
which lies in each cell. The space of cells is assumed to be homogeneous enough in order 
to ensure that each cell has the same number of neighbours. This condition is naturally 
satisfied if the space of cells is associated to a tiling which is a tessellation based
on a single regular tile. Then, each cell is associated to a tile which is called the
{\bf support} of the cell. Each cell has a state belonging
to some finite set~$\cal L$, called the set of states. As $\cal L$ is finite, it can
be seen as the alphabet used by the finite automaton which equips the cells. 
The cellular automaton evolves in a discrete time provided by a clock. 
At time~$t$, each cell updates its state according to the current value of its state at
time~$t$ and the values at the same time of the states of its neighbours. These 
current states constitute the neighbourhood to which the finite automaton associates a
new state which will be the current state of the cell at time~$t$+1. The way with which 
such an association is performed is called a {\bf rule} of the cellular automaton.
There are finitely many rules constituting the {\bf program} of the cellular automaton.

   A {\bf quiescent} state is a state~$\xi$ such that the cell remains in state~$\xi$
if all its neighbours are also in state~$\xi$. The corresponding rule is called the
{\bf quiescent rule}. Usually, we call that state {\bf white}
and it is denoted by~{\tt W}. A {\bf configuration} at time~$t$ is the set of cells 
which are in a non-quiescent state together with the position of their supports in the 
tiling. Traditionally, the {\bf initial} configuration of a cellular automaton
is {\it finite}. This means that at time~0, the time which marks the beginning of the
computation, the set of cells which are in the non-quiescent state is finite.
Define the {\bf distance} of a cell~$c$ to a cell~$d$ by the smallest number of cells needed
to link~$c$ to~$d$ by a sequence of cells where two consecutive ones are neighbours.
Then, define the {\bf disc} $D(c,n)$ of center~$c$ and radius~$n$ as the set of cells~$d$
whose distance from~$c$ is at most~$n$. If we fix a cell~$c$ as origin of the space,
there is a smallest number~$N_0$ such that the initial configuration is contained in
$D(c,N_0)$. This means that all cells outside $D(c,N_0)$ are in the quiescent state.
Call such an $N_0$ the {\bf initial border number}. The reason of the index~0 will be clear 
later.
Let $C(c,n)$ be the set of cells whose distance from~$c$ is exactly~$n$. The definition
of~$N_0$ also entails that $C(c,N_0)$ contains at least one non-quiescent state.
In this setting, the halting of a cellular automaton is reached 
by two identical consecutive configurations. Accordingly, there is a number~$k$ and a 
time~$t$ such that the configurations at time~$t$ and $t$+1 are both contained in
$D(c,k)$ and they are equal. 

From now on, when we say cellular automaton, 
we need to understand deterministic cellular automaton with a quiescent state. The term 
{\bf deterministic} means that a unique new state is associated to the current state of
a cell and the current states of its neighbours.

   From various papers of the author, we know the following on cellular automata in
hyperbolic spaces: in the tessellations $\{5,4\}$, $\{7,3\}$ and $\{5,3,4\}$, namely
the pentagrid, the heptagrid and the dodecagrid respectively, it is possible to construct
weakly universal cellular automata with two states only. In the case of the dodecagrid,
the constructed automaton is rotation invariant, we remind the definition in
Section~\ref{sdecidri}. 
In the case of the pentagrid and of the
heptagrid, the rules are not rotation invariant. Moreover, in the case of the pentagrid,
we assume the Moore neighbourhood, {\it i.e.} we assume that the neighbours of the
cell are the cells which share at least a vertex with it. It is known that with rotation
invariant rules and von Neumann neighbourhood, which means that the neighbours of a cell
share a side with it, there is a strongly universal cellular automaton in the pentagrid
with ten states, see \cite{mmSU}. This means that the cellular automaton which is universal
starts its computation from a finite configuration. If we relax the rotation 
invariance, there is a weakly universal cellular automaton in the pentagrid with
five states. And so, results concerning rotation invariance are also interesting.

    Very little is known if we change something in the above assumptions.

    The present paper is devoted to the proof of our main result:

\begin{thm}\label{decid}
For any deterministic cellular automaton in the pentagrid, if its initial configuration 
is finite and if it has at most two states with one of them being quiescent, 
then its halting problem is decidable.
\end{thm}

The proof is split into two propositions dealing first with rotation invariance
in Section~\ref{sdecidri}, then when that condition is relaxed, see 
Section~\ref{sdecidnri}. In Section~\ref{sfront} we study what happens in an infinite
motion of the cellular automaton when such a motion occurs. In Section~\ref{pentagrid}
we present to the reader a minimal introduction of the pentagrid and of the implementation
of cellular automata in that context. Section~\ref{conclude} brings in a few reflections
on the topic.

We now turn to hyperbolic geometry and the tiling we consider in which the cellular automata
later considered evolves.

\section{The pentagrid}\label{pentagrid}

   In this paper, we use the model of the hyperbolic geometry which is known as
the {\it Poincar\'e's disc}. A disc is fixed, call it the unit disc. Let $D$ be the open
unit disc. The model~$\cal M$ of the hyperbolic plane we consider is defined in $D$
which we call the {\bf support} of $\cal M$. 
The points in~$\cal M$ are the points of the open disc. The lines in~$\cal M$ are the
traces in~$D$ of circles which are orthogonal to $\partial D$, the border of~$D$ and the
traces in~$D$ of straight lines which pass through the centre of~$D$. 

Figure~\ref{poincar} represents a line~$\ell$ and a point~$A$ out of~$\ell$. The
figure also shows us four lines which pass through~$A$. The line~$s$ cuts $\ell$ and
is therefore called a {\bf secant} with~$\ell$. The lines~$p$ and~$q$ touch~$\ell$
on~$\partial D$. The points $P$ and $Q$ where, respectively, $p$ and $q$ touch~$\ell$
are called {\bf points at infinity} of the hyperbolic plane but do not belong to that
plane. The lines~$p$ and~$q$
are called {\bf parallel} to~$\ell$. At last, but not the least, the line~$m$ does
not cut~$\ell$ and it also does not touch it neither in~$D$ nor on its border, nor 
outside~$D$. The line~$m$ is called {\bf non-secant} with~$\ell$. It is proved that
two lines of the hyperbolic plane are non-secant if and only if they have a unique common
perpendicular.

\vskip 10pt
\vtop{
\ligne{\hfill
\includegraphics[scale=1]{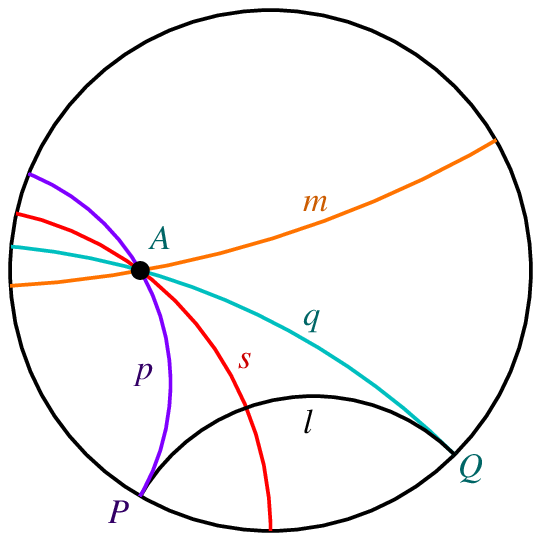}\hfill}
\begin{fig}\label{poincar}
\leurre
Poincar\'e's disc model of the hyperbolic plane. Here, the various relations between
a line, a point out of the line with other lines passing through the point.
\end{fig}
}

   A theorem by Poincar\'e tells us that there are infinitely many tessellations in the 
hyperbolic plane whose basic tile is a triangle with angles 
$\displaystyle{\pi\over p}$,
$\displaystyle{\pi\over q}$ and
$\displaystyle{\pi\over r}$ provided that the positive numbers $p$, $q$ and~$r$
satisfy
\vskip 5pt
\ligne{\hfill
$\displaystyle{\pi\over p} + \displaystyle{\pi\over q} + \displaystyle{\pi\over r}
< 1$,
\hfill}
\vskip 5pt

\noindent
which simply means that the triangle with these angles lives in the hyperbolic plane.
As a consequence, if we consider $P$ the regular convex polygon with $p$ sides and
with interior angle $\displaystyle{{2\pi}\over q}$, $P$ tiles the plane by recursive
reflections in its sides and in the sides of its images if and only if
\vskip 5pt
\ligne{\hfill
$\displaystyle{\pi\over p} + \displaystyle{\pi\over q} < \displaystyle{1\over 2}$.
\hfill}
\vskip 5pt
When this is the case, the corresponding tessellation is denoted by $\{p,q\}$.

We call {\bf pentagrid} the tessellation $\{5,4\}$ which is illustrated by
Figure~\ref{penta}.

In~\cite{mmJUCSii,mmbook1}, it is proved that the pentagrid is spanned by a tree.
The left hand-side of Figure~\ref{quarter} shows us a quarter of the pentagrid
spanned by the tree illustrated on the right-hand side of the same figure.
That tree is called the {\bf Fibonacci tree}. The reason of this name comes
from the properties of the tree. The tree is a finitely branched tree generated
by two rules: 
\vskip 5pt
\ligne{\hfill
$B\longrightarrow BW$ and $W\longrightarrow BWW$.
\hfill$(R_F)$\hskip 10pt}
\vskip 10pt
\vtop{
\ligne{\hfill
\includegraphics[scale=0.8]{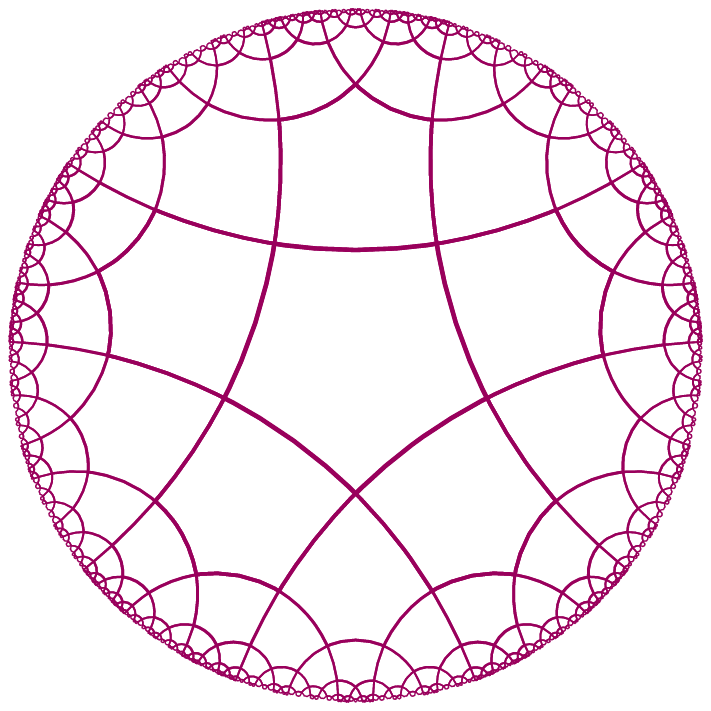}\hfill}
\begin{fig}\label{penta}
\leurre
The pentagrid as it can be represented in Poincar\'e's disc model of the hyperbolic 
plane. 
\end{fig}
}
\vskip 5pt
Indeed, we split the nodes into two kinds: black nodes and white nodes. Black nodes
have two sons, as suggested by the above rules, a black son, the left-hand side son,
and a white son, the right-hand side one. White nodes have three sons, a black son,
the leftmost one, and two white sons, the others. The root of the tree is a white node.
It is not difficult to prove, see~\cite{mmJUCSii,mmbook1} from the above rules that
there are exactly $f_{2n+1}$ nodes lying on the $n^{\rm th}$ level of the tree, where
$f_k$ is the Fibonacci sequence where \hbox{$f_0=f_1=1$}. 

There is another, more striking
property. Number the nodes of the tree, starting from the root, which receives~1, and
level after level and, on each level from left to right. Then, represent these numbers
in the Fibonacci sequence, choosing the one whose number of digits is the biggest.
Then, if $[n]$ is that Fibonacci representation of~$n$, the black son of the node~$n$
has the number represented by $[n00]$ if $n$ is attached to a black node and
the middle son of the node~$n$ has the number with the same representation if $n$ is
attached to a white node. This property is called the {\bf preferred son} property which
can be checked on the right-hand side picture of Figure~\ref{quarter}.

\vskip 10pt
\vtop{
\ligne{\hfill
\includegraphics[scale=1]{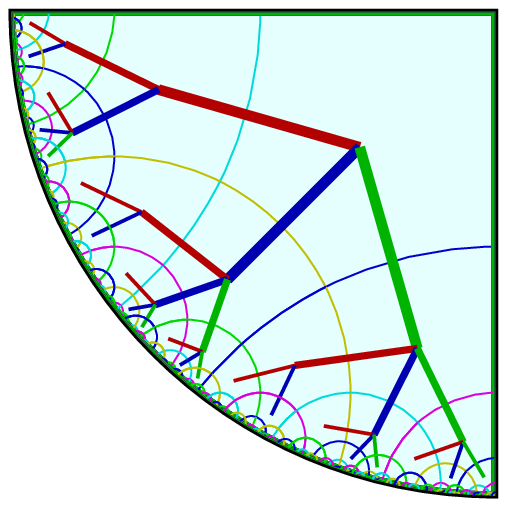}\hfill
\includegraphics[scale=1]{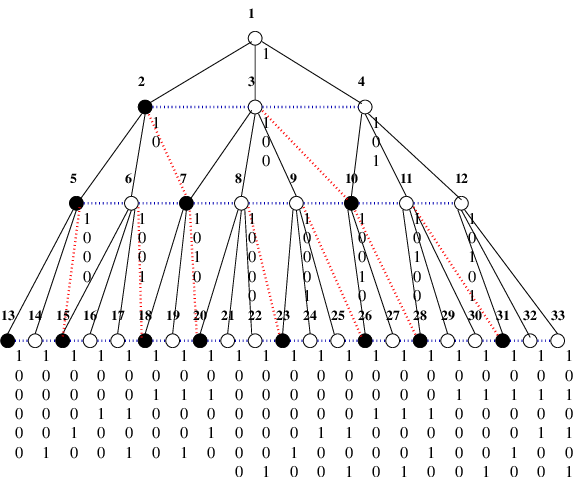}\hfill}
\begin{fig}\label{quarter}
\leurre
To left: A sector of the pentagrid generated by the Fibonacci tree illustrated to right.
In the right-hand side picture: under each node, vertically, we represented the Fibonacci
representation of the number attached to the node. We can check the preferred son 
property.
\end{fig}
}

   In the sequel, we shall also call Fibonacci tree a tree whose root is a black node
and to which the rules~$(R_F)$ are applied to its sons and, recursively, to the sons of
its sons. In such a Fibonacci tree, the number of nodes on the $n^{\rm th}$ level of the
tree is $f_{2n}$.

   Note that in Figure~\ref{penta}, a tile seems to play a different role than the others.
It is the tile which contains the centre of the support of $D$. As can be seen
in Figure~\ref{penta}, not much can be seen from the tiling. We can very well see the
central tile, also well its neighbours, but going further from the central tile, we can
see the tiles less and less. In fact, as the hyperbolic plane has no centre, the pentagrid
too has no tile playing a central role. We can view the support of our model as a 
{\it window} over the hyperbolic plane. We can imagine that we fly over that plane,
that the window is a screen on the board of our space craft. The centre of that window
is simply the point of the hyperbolic plane over which our space craft is flying. 
Indeed, we fly with instruments only, those which we just defined. This window property
of the Poincar\'e's disc stresses that so little can be represented of this space 
in its Euclidean models. It is the reason why we choose the disc model.

We fix a tile~$\tau_0$ which we call from now on the 
{\bf central tile} and we shall consider that the central tile is the tile in which the 
centre of~$D$ lies in the figures.
As illustrated by the left-hand side of Figure~\ref{fibo_disc}, around the central tile, 
we can assemble five quarters as those defined in the left-hand side picture of 
Figure~\ref{quarter} in order to construct the whole pentagrid. We call these quarters 
{\bf sectors}. In each sector, the tiling is spanned by the Fibonacci tree. 
It is not difficult to prove that the tiles which lie on the level~$k$ of a Fibonacci tree 
of a sector are at the distance~$k$ from the central tile. We call {\bf Fibonacci circle 
of level~$n$} the set of tiles $C(\tau_0,n)$ denoted by~${\cal F}_n$. Similarly, we call 
{\bf Fibonacci discs} the sets $D(\tau_0,n)$ which we denote by ${\cal D}_n$.
Note that ${\cal D}_n$ is the union of the Fibonacci circles ${\cal F}_k$ with 
\hbox{$0\leq k\leq n$}. In Figure~\ref{fibo_disc}, we illustrate the notion of Fibonacci 
circles and discs by marking in blue, green and gray the tiles which belong to
${\cal F}_3$ and by marking in pink those which belong to ${\cal D}_2$.

   We call the Fibonacci representation we attached to the number given to a node~$\nu$ of
a Fibonacci tree the {\bf coordinate} of~$\nu$ denoted by $[\nu]$. We identify the node 
with its number~$\nu$ and sometimes by~$[\nu]$. We locate the tiles of the 
pentagrid with~0 for the central tile and for the other tiles with 
two numbers: the number of the sector in which the tile lies and the number of the
node in the Fibonacci tree which spans the sector, as clear from Figure~\ref{fibo_disc}.
We extend the coordinate of a tile by appending the number of its sector.
We shall also say that the central tile is the support of the 
{\bf central cell}. Again, the central cell is the cell on which we focus our attention
at the given moment of our argumentation.
\vskip 10pt
\vtop{
\ligne{\hfill
\includegraphics[scale=0.9]{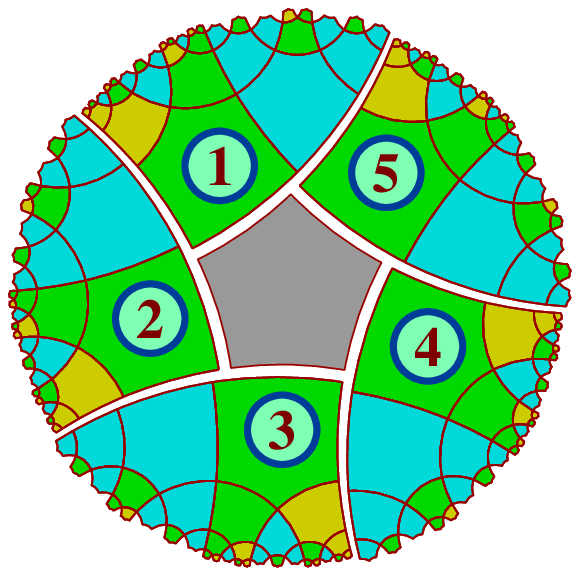}\hfill
\includegraphics[scale=0.72]{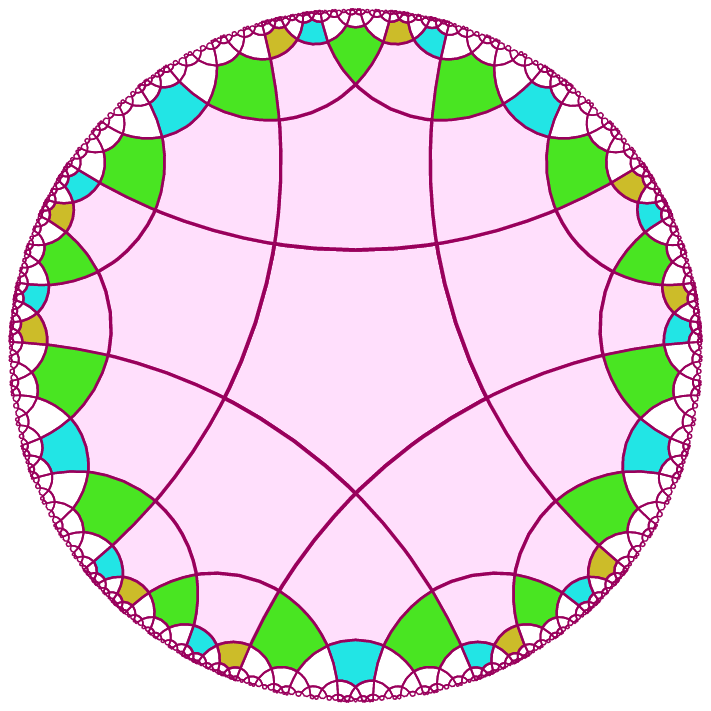}\hfill}
\begin{fig}\label{fibo_disc}
\leurre
To left, how sectors are assembled around the central cell in order to get the
pentagrid. To right, the Fibonacci circle of level~$3$. Together with the tiles of the 
Fibonacci circle, the tiles in pink, i.e. the central cell and the tiles of levels~$0$ 
and~$1$ constitute the Fibonacci disc of level~$2$.
\end{fig}
}

These considerations allow us to implement cellular automata in
the pentagrid as performed in~\cite{mmJUCSii,mmbook2}. As mentioned in the introduction,
to each tile we associate a cell of the cellular automaton. We shall also identify the
cell by the coordinates of its support, or its number depending on the context. 
If $\eta$ is the state of the cell attached to the tile~$\nu$, we say that $\nu$ is
also an $\eta$-cell. In order to note the rules of a deterministic cellular automaton in 
the pentagrid, we introduce a numbering of the sides of each tile. The numbering
starts from~1 and it is increased by~1 for each side while counterclockwise turning around
the tile. For the central cell, 
side~1 is fixed once and for all and for the other tiles, side~1 is the side of the tile 
shared with its father, the central cell being the father of the root of the tree. 
Neighbour~$i$ of a cell~$\nu$ shares with~$\nu$ the side~$i$ of~$\nu$.
The precision is required because the side shared by two tiles do not receive the same 
number in both tiles.

If $\eta_0$ is the current state of the cell, if $\eta_0^1$ is its new state and if
$\eta_i$ is the state of its neighbour~$i$ at the current time, then the rule
giving $\eta_0^1$ from $\eta_0$ and the $\eta_i$'s is written as a word in 
\hbox{$\{\cal L\}^\ast$}, where $\cal L$ is the set of states of the cellular automaton:
\hbox{$\underline{\eta_0}\eta_1...\eta_5\underline{\eta_0^1}$}. The underscore is put 
under $\eta_0$ and~$\eta_0^1$ in order to facilitate the reading.
In a rule \hbox{$\underline{\eta_0}\eta_1...\eta_5\underline{\eta_0^1}$}, we say that 
\hbox{$\underline{\eta_0}\eta_1...\eta_5$} is the {\bf context} of the rule
and we say that the word \hbox{$\eta_1...\eta_5$} is the {\bf state neighbourhood} of 
the cell. 

\section{Rotation invariant cellular automata in the pentagrid
with two states}\label{sdecidri}

By definition, the rules of a cellular automaton~$A$ in the pentagrid are said to be 
{\bf invariant by rotation}, in short {\bf rotation invariant} and $A$ is said to
be a {\bf rotation invariant cellular automaton},
if for each rule present in the program of~$A$, namely,
\hbox{$\underline{\eta_0}\eta_1...\eta_5\underline{\eta_0^1}$}, the rules
\hbox{$\underline{\eta_0}\eta_{\pi(1)}...\eta_{\pi(5)}\underline{\eta_0^1}$} 
are also present, where $\pi$ runs over the circular permutations on \hbox{$[1..5]$}. 
When the
cellular automaton is rotation invariant, we usually indicate the rule where, after
the current state, we have the state of neighbour~1.

The goal of this section is to prove

\begin{prop}\label{sdecid2ri}
For any deterministic cellular automaton in the pentagrid, if its 
initial configuration 
is finite, if it has two states with one of them being quiescent, and if its rules are 
rotation invariant, then its halting problem is decidable.
\end{prop}

   Our proof is based on the following considerations. If the halting of the computation
of a cellular automaton halts, it means that the computation remains in some ${\cal D}_N$
for ever. Note that the computation may remain within some ${\cal D}_N$ and not halt. But
in that case, after a certain time, the computation becomes periodic. And this can be 
detected: it is enough to find two identical configurations during the computation: this
generalizes the situation of the halting. What is not that easy to detect is the case
when the configuration extends to infinity in that sense that for each circle ${\cal F}_k$,
there is a time when that circle contains a non quiescent cell.

Let us closer look at such a case.
Let $N_0$ be the initial border number. We know that there is at 
least one
tile~$\nu$ of~${\cal F}_{N_0}$ which is a {\tt B}-cell, at time~0. Call ${\cal F}_{N_0}$ the
{\bf front} at time~0. The front at time~$t$ is ${\cal F}_{N_t}$ where $N_t$~is the 
smallest~$k$ such that ${\cal F}_k$ contains all configurations at time~$\tau$, 
with \hbox{$\tau\leq t$}, and such that 
all cells outside ${\cal D}_k$ are quiescent. This is the reason why the initial border 
number is denoted by~$N_0$: ${\cal F}_{N_0}$ is the front at time~0. 

Our proof of Proposition~\ref{sdecid2ri} lies on the analysis of how a {\tt B}-cell
on the front at time~$t$ can propagate to the front at time~$t$+1. If we can prove
that $N_t$ is a non-decreasing function of~$t$ which tends to infinity, we then prove
that the computation of the cellular automaton does not halt. The main property
which will allow us to detect such a situation is that a cell on ${\cal F}_{n+1}$,
has at most two neighbours on ${\cal F}_n$ and the others on ${\cal F}_{n+2}$. 
So that if $\nu$ is node of the front which is a {\tt B}-cell, the state neighbourhood
of its sons is either {\tt BW$^4$} or {\tt B$^2$W$^3$}. That situation occurs if and only
if the node~$\nu$$-$1 of the front is also a {\tt B}-cell. We say that a {\tt B}-cell
is {\bf isolated} on ${\cal F}_n$ if $\nu$ being its support, $\nu$$-$1 and $\nu$+1 are
both {\tt W}-cells. These considerations significantly reduce the number of rules to 
consider and, consequently, the number of cases to scrutinize.
More precisely, we have the following lemma.

\begin{lem}\label{front}
Let $A$ be a deterministic cellular automaton on the pentagrid with two states, one being
quiescent, and whose rules are rotation invariant. If the rule
\hbox{\tt$\underline{\hbox{\tt W}}$BW$^4\underline{\hbox{\tt W}}$} 
occurs in the program of~$A$, the front at time~$t$$+$$k$ is the same as the front at 
time~$t$$+$$1$ for \hbox{$k\geq2$}, {\it i.e.} \hbox{$N_t$$+$$k = N_t$$+$$1$} for the 
same values of~$k$.  If it is not the case, {\it i.e.} if the rule 
\hbox{\tt$\underline{\hbox{\tt W}}$BW$^4\underline{\hbox{\tt B}}$} 
occurs in the program of~$A$, then if the front at time~$t$ contains
a {\tt B}-cell, the front at time~$t$$+$$1$ also contains a {\tt B}-cell, {\it i.e.} we 
have $N_{t+1} = N_t$$+$$1$.
\end{lem}

\noindent
Proof of the lemma. Let~$\nu$ be the tile of~${\cal F}_{N_t}$ which is a {\tt B}-cell.
Assume that $\nu$ is an isolated {\tt B}-cell of the front at time~$t$. Let $\sigma$ be 
a son of~$\nu$. Whether $\sigma$ is a black node or a white one, $\sigma$ is a {\tt W}-cell
as well as its sons. Accordingly, its state neighbourhood is {\tt BW$^4$} so that
the rule
\hbox{\tt$\underline{\hbox{\tt W}}$BW$^4\underline{\hbox{\tt W}}$} applies. Consequently,
$\sigma$ remains a {\tt W}-cell at time~$t$.

If $\nu$+1 is also a {\tt B}-cell at time~$t$. Let $\sigma$ be the black son of~$\nu$+1.
Then, the state neighbourhood of~$nu$ is {\tt B$^2$W$^3$}. If the program of~$A$
contains the rule
\hbox{\tt$\underline{\hbox{\tt W}}$B$^2$W$^3\underline{\hbox{\tt W}}$}, then
$\sigma$ remains a {\tt W}-cell at time~$t$+1 as well as the other sons of~$\nu$. 
If the program contains the rule
\hbox{\tt$\underline{\hbox{\tt W}}$B$^2$W$^3\underline{\hbox{\tt B}}$}, then
$\sigma$ becomes a {\tt B}-cell at time~$t$+1 but the cells~$\sigma$$-$1 and $\sigma$+1 
are white nodes, so that whatever the state of their father, they remain {\tt W}-cells
at time~$t$+1 as either the quiescent rule or the rule
\hbox{\tt$\underline{\hbox{\tt W}}$BW$^4\underline{\hbox{\tt W}}$} 
applies to them. Accordingly, in that case, the cell~$\sigma$ is an isolated {\tt B}-cell
of the front at time~$t$+1. Now, from what we proved in the previous paragraph shows us
that the sons of~$\sigma$ remain {\tt W}-cells at the time~$t$+1 so that the front
at time~$t$+2 is the same as at time~$t$+1 and it remains the same afterwards. This proves
the part of the lemma concerning the rule
\hbox{\tt$\underline{\hbox{\tt W}}$BW$^4\underline{\hbox{\tt W}}$}.

Now,
assume that the rule
\hbox{\tt$\underline{\hbox{\tt W}}$BW$^4\underline{\hbox{\tt B}}$} occurs in the program
of~$A$. From our previous study on the sons of~$\nu$, at least one of them is a white
node which means that its neighbourhood is {\tt BW$^4$}. Accordingly, if $\nu$ is a 
{\tt B}-cell, that white son becomes a {\tt B}-cell at the next time, so that
\hbox{${\cal F}_{N_{t+1}} = {\cal F}_{N_t+1}$}.\hfill$\Box$

We are now in position to prove Proposition~\ref{sdecid2ri}. 
If the initial configuration is empty, {\it i.e.} if all tiles are {\tt W}-cells at time~0,
there is nothing to prove: the configuration remains empty for ever. Accordingly, if
the initial configuration is not empty,
$N$ is definite, so that ${\cal F}_N$ contains at least one {\tt B}-cell.
From Lemma~\ref{front}, if the rule
\hbox{\tt$\underline{\hbox{\tt W}}$BW$^4\underline{\hbox{\tt B}}$} occurs in the program
of~$A$, the front moves by one step forward at each time, so that the computation of
the cellular automaton does not halt. If that rule does not occur then, necessarily,
the rule
\hbox{\tt$\underline{\hbox{\tt W}}$BW$^4\underline{\hbox{\tt W}}$} is present in the 
program of~$A$. From Lemma~\ref{front}, we know that at most, we have
\hbox{${\cal F}_{t_1} = {\cal F}_1$} but that necessarily, 
\hbox{${\cal F}_{t_k} = {\cal F}_1$} for \hbox{$k\geq1$}.\hfill$\Box$

\section{When the rules are not rotation invariant}\label{sdecidnri}

Here again, we deal with a deterministic cellular automaton with a quiescent state 
which starts its computation from a finite configuration. But in this section, we relax
the assumption of rotation invariance. The convention we fixed in Section~\ref{pentagrid}
for the numbering of the sides of a tile have their full meaning in this section.
And so, a rule 
\hbox{$\underline{\eta_0}\eta_1...\eta_5\underline{\eta_0^1}$} may be 
different from a rule
\hbox{$\underline{\eta_0}\eta_{\pi(1)}...\eta_{\pi(5)}\underline{\eta_0^1}$} 
where $\pi$ is a permutation over \hbox{$[1..5]$}.
Note that this time, the order of
the letters in the state neighbourhood associated to the rule is meaningful. 

   Consider a cell~$\nu\in{\cal F}_{n+1}$. In all cases, its
neighbour~1 is its father which by construction belongs to~${\cal F}_n$. If $\nu$ is a
black node, as already noticed in previous sections, $\nu$ has two neighbours exactly
which belong to~${\cal F}_n$: neighbour~1, as it is the father and also neighbour~2.
Consider $N_0$ the initial border number. From what we just noticed, a rule can 
make a state~{\tt B} move 
from ${\cal F}_{N_0}$ to ${\cal F}_{N_0+1}$ if its state neighbourhood starts with
{\tt  BW}, {\tt  WB} or {\tt B$^2$}: the last two cases may happen if the considered 
cell of~${\cal F}_{N_0+1}$ is a black node. As an example, the state neighbourhood of 
the tile~$\nu$ of~${\cal F}_{N_0+1}$ cannot be {\tt WWBWW}: if a rule whose state
neighbourhood is {\tt WWBWW} is applied to a cell of~${\cal F}_{n}$, its neighbour
which is a {\tt B}-cell belongs to ${\cal F}_{n+1}$. 

\begin{lem}\label{frontnri}
Let $A$ be a deterministic cellular automaton on the pentagrid with two states where one
of them is a quiescent state. If the program of $A$ contains the rule
\hbox{$\underline{\tt W}$\tt BW$^4\underline{\hbox{\tt W}}$} and the rule 
\hbox{$\underline{\tt W}$\tt WBW$^3\underline{\hbox{\tt W}}$}, 
then ${\cal F}_{N_1+k} = {\cal F}_{N_1}$ for all positive integer $k$ with
\hbox{$k\geq2$}.
\end{lem}

\noindent
Proof of the lemma. The proof comes from the fact that the state neighbourhood
of a the son of a node~$\nu$ which is an isolated {\tt B}-cell of the front is 
{\tt BW$^4$}. If $\nu$ is a 
{\tt B}-cell and if $\nu$+1 is a {\tt W}-cell, then the state neighbourhood of the
leftmost son of~$\nu$+1 is {\tt WBW$^3$}. Now,
the sons of an isolated {\tt B}-cell of the front at time~$t$ remain quiescent at 
time~$t$+1.
If the program of~$A$ contains the rule
\hbox{$\underline{\tt W}$\tt B$^2$W$^3\underline{\hbox{\tt B}}$}, then
if the state pattern {\tt BB} is present on the front at time~$t$, say at the nodes~$\nu$
and~$\nu$+1, then the just mentioned rule applies to the leftmost son $\sigma$ of
$\nu$+1, but the white sons of~$\nu$ and those of~$\nu$+1 remains {\tt W}-cells at
time~$N_t$+2. Accordingly, $\sigma$ is an isolated {\tt B}-node of the level $N_t$+1
so that, from the just previous study, all cells on ${\cal F}_{N_t+2}$ remain
quiescent, so that ${\cal F}_{N_t+k} = {\cal F}_{N_t+2}$ for all positive integer~$k$.
Clearly, the same conclusion holds if the program of~$A$ contains the rule
\hbox{$\underline{\tt W}$\tt B$^2$W$^3\underline{\hbox{\tt W}}$}.\hfill$\Box$

Let $\nu_1$, ..., $\nu$+$k$ a sequence of consecutive nodes on the circle~${\cal F}_n$.
Then, the word $\eta_1..\eta_k$ with \hbox{\tt $\eta_i\in\{$B$,$W$\}$}, 
\hbox{$i\in\{1..k\}$}, is called a {\bf state pattern}.

\newdimen\LLlarge \LLlarge=30pt
\def\Lmligne #1 #2 #3 #4 #5 #6 #7 {
\ligne{%
\hbox to \LLlarge {\hfill#1}
\hbox to \LLlarge {\hfill#2}
\hbox to \LLlarge {\hfill#3}
\hbox to \LLlarge {\hfill#4}
\hbox to \LLlarge {\hfill#5}
\hbox to \LLlarge {\hfill#6}
\hbox to \LLlarge {\hfill#7}
\hfill}
}
\newdimen\TRlarge \TRlarge=60pt
\def\TRligne #1 #2 #3 #4 {
\ligne{%
\hbox to \LLlarge{\hfill#1\hfill}
\hbox to \TRlarge{\hfill#2\hfill}
\hbox to \LLlarge{\hfill#3\hfill}
\hbox to \TRlarge{\hfill#4\hfill}
\hfill}
}
\def\ab{\hbox{\footnotesize\tt BW}}
\def\wb{\hbox{\footnotesize\tt WB}}
\def\bb{\hbox{\footnotesize\tt BB}}
\def\abw{$\overline{\hbox{\footnotesize\tt BW}}$}
\def\wbw{$\overline{\hbox{\footnotesize\tt WB}}$}
\def\bbw{$\overline{\hbox{\footnotesize\tt BB}}$}
\ligne{\hfill
\vtop{\leftskip 0pt\parindent 0pt\hsize=300pt
\begin{tab}\label{Tfront}
\leurre
Rules of a deterministic cellular automaton on the pentagrid with two states,
{\tt W} being the quiescent state, which apply to the sons of a node of the front. 
\end{tab}
\ligne{\hfill
\vtop{\leftskip 0pt\parindent 0pt\hsize=200pt
\TRligne {\ab} 
{{\tt$\underline{\hbox{\tt W}}$BW$^4\underline{\hbox{\tt B}}$}}
{\abw}
{{\tt$\underline{\hbox{\tt W}}$BW$^4\underline{\hbox{\tt W}}$}}
\TRligne {\wb} 
{{\tt$\underline{\hbox{\tt W}}$WBW$^3\underline{\hbox{\tt B}}$}}
{\wbw}
{{\tt$\underline{\hbox{\tt W}}$WBW$^3\underline{\hbox{\tt W}}$}}
\TRligne {\bb} 
{{\tt$\underline{\hbox{\tt W}}$B$^2$W$^3\underline{\hbox{\tt B}}$}}
{\bbw}
{{\tt$\underline{\hbox{\tt W}}$B$^2$W$^3\underline{\hbox{\tt W}}$}}
}
\hfill}
}
\hfill}
\vskip 10pt

Let us now prove Theorem~\ref{decid}.
From Lemma~\ref{frontnri}, the computation remains within ${\cal D}_{N_1+2}$ if
the program of~$A$ contains both rules \abw{} and \wbw. 
Accordingly, we may assume that it contains either the rule \ab{} or the rule \wb.

Consider the case when the rule \ab{} 
belongs to the program of~$A$. 
If a node~$\nu$ of the front at time~$t$ is a {\tt B}-cell, from the proof of 
Lemma~\ref{frontnri} we know that there is also a {\tt B}-cell on the front at time~$t$+1 
and that we have \hbox{$N_{t+1} = N_t$+1} as any white son of~$\nu$ is a {\tt B}-cell 
on ${\cal F}_{N_t+1}$ at time~$N_t$+1.

Consider the case when the rules \abw{} and \wb{}
belong to the program of~$A$. If a node~$\nu$ of the front
at time~$t$ is a {\tt B}-cell we have to look at the case when the state pattern
{\tt BW} is present on the front or not. As by definition the front contains at least
one {\tt B}-cell, if the state pattern {\tt BW} is not present, this means that all tiles
of the front are {\tt B}-cells. In that case, all black nodes of ${\cal F}_{N_t+1}$ have
the state neighbour {\tt B$^2$W$^3$}. Accordingly, as the white nodes of ${\cal F}_{N_t+1}$
remain quiescent at time~$t$+1, the evolution depends on the rule whose context is
\hbox{\tt$\underline{\hbox{\tt W}}$B$^2$W$^3$}. 

If the rule is \bbw, 
then the black nodes
of ${\cal F}_{N_t+1}$ remain quiescent at time~$t$+1, which entails that all nodes of
${\cal F}_{N_t+1}$ remain quiescent at time~$t$+1. Now, if at time~$t$+1 at least one 
node of ${\cal F}_{N_t}$ at time~$t$ becomes a {\tt W}-cell at time~$t$+1 and at least 
one remains a {\tt B}-cell, then the pattern {\tt BW} occurs, say on the nodes $\nu$ and 
$\nu$+1. Then, if $\sigma$ is the black node of~$\nu$+1, its state neighbourhood at 
time~$t$+1 is {\tt WBW$^3$}, so that the rule \wb{}
applies and $\sigma$ becomes
a {\tt B}-cell at time~$t$+2. From the rule \abw,
we know that the white sons
of the {\tt B}-cells on ${\cal F}_{N_t}$ remain {\tt W}-cells and so the black nodes
on ${\cal F}_{N_t+1}$ are isolated {\tt B}-cells on the level $N_t$+1. Accordingly,
the rule \wb{}
applies to their
black sons on ${\cal F}_{N_t+2}$ at time~$t$+2. The argument applies again to those nodes
which are also isolated {\tt B}-cells on the new front. So that
${\cal F}_{N_{t+k}} = {\cal F}_{N_t+k}$ for all positive integer~$k$.

We remain with the situation when all the nodes of the front at time~$t$ are {\tt B}-cells
and all of them become {\tt W}-cells at time~$t$+1. We can repeat the above analysis
to time~$t$+2. If at that time all nodes are again {\tt B}-cells, say that this situation 
is an alternation of {\tt B} and {\tt W}. If such a situation is repeated long enough, as
in that case the front does not go beyond ${\cal D}_{N_t+1}$, the computation remains
for ever within that disc and so the computation is periodic. We know that such an
evolution can be detected: it is enough to observe two identical configurations. If this
is not the case, we find a situation where the front contains the state pattern {\tt BW}
so that the rule \wb{}
applies endlessly as already
seen.

If the program of~$A$ contains the rule \bb,
as it also contains
the rule \abw,
the state pattern
{\tt BW} occurs on ${\cal F}_{N_t+1}$ at time~$t$+1 as soon as the pattern {\tt BB} 
occurs on ${\cal F}_{N_t}$ at time~$t$. If the pattern {\tt BB} does not occur, 
clearly, the pattern {\tt BW} occurs on ${\cal F}_{N_t}$ at time~$t$, so that
we have the same conclusion as we had with the rule \bbw{} when the pattern {\tt BW}
occurs on the front: a non-halting computation which is detected by the occurrence
of that pattern.

We can summarize the discussion as follows:
\newdimen\EClarge \EClarge=70pt
\newdimen\EClargebis \EClargebis=100pt
\def\ECligne #1 #2 #3 {
\ligne{
\hbox to \TRlarge{\hfill#1\hfill}
\hbox to \EClarge{\hfill#2\hfill}
\hbox to \EClargebis{\hfill#3\hfill}
\hfill}
}
\vskip -5pt
\ligne{\hfill
\vtop{\leftskip 0pt\parindent 0pt\hsize=350pt
\begin{tab}\label{Tdecid}
\leurre
Table of the evolutions of the computation of~$A$ depending on its rules of 
Table~{\rm\ref{Tfront}} and on the patterns which can be seen on the front.
\end{tab}
\ligne{\hfill
\vtop{\leftskip 0pt\parindent 0pt\hsize=290pt
\ECligne {rules} {front} {evolution}
\ECligne {\ab} {a {\tt B}-cell} {$N_{t+1} = N_t$$+$$1$ from some~$t_0$}
\ECligne {\abw, \wbw} {any} {within ${\cal D}_{N_{t_0}}$ from some~$t_0$}
\ECligne {\abw, \wb} {a {\tt BW}} {$N_{t+1} = N_t$$+$$1$ from some~$t_0$}
\ECligne {\abw, \wb} {never {\tt BW}} {within ${\cal D}_{N_{t_0}}$ from some~$t_0$}
}
\hfill}
}
\hfill}
\vskip 10pt

Accordingly, as we have analyzed all possible cases each of one can be detected,
we conclude that the proof of Theorem~\ref{decid} is completed.\hfill$\Box$

\section{Propagation of the front}
\label{sfront}

   Although we solved the question about the halting problem for such cellular automata,
it can be interesting to examine their behaviour in the case when the computation does
not halt with an unbounded occurrence of non quiescent cells. We shall focus on the
front. Up to now, we have seen that a motion to infinity exactly means that the front
is increasing starting from some time~$t_0$. This happens in different settings
as shown by Table~\ref{Tdecid}. It could be interesting to have more information about
such a motion. However, as the situation may be intricate in some cases as can be seen
in the proof of Theorem~\ref{decid} when the rules are \abw, \wb{} and \bbw, we
shall restrict our attention to what happens on the front.
We shall see that with two states only the study of this restricted aspect is 
not that trivial.

   From Table~\ref{Tdecid}, we know that we basically have to consider two cases:
the case when the program of~$A$ contains the rule \ab{} and the case when it contains
the rule \abw{} together with the rule~\wb{} in the case that the pattern {\tt BW} appears
at some time on the front.

Consider that latter case. From the proof of Theorem~\ref{decid}, we know that if
the pattern {\tt BW} appears on the front at time~$t$, it will also appear on the front
at time~$t$+1. But the proof has given us a more exact information. If $\nu$ is the node
of the {\tt B}-cell of some {\tt BW} pattern of the front at time~$t$, the application of 
the rule \wb{} to the black son~$\sigma$ of~$\nu$+1 produces a {\tt BW} patterns on the 
nodes $\sigma$ and $\sigma$+1 as $\sigma$+1 remains a quiescent cell due to the fact 
that $\nu$+1 is a {\tt W}-cell and that $\sigma$+1 is a white node. And so,
$\sigma$ and $\sigma$+1 define a {\tt BW} pattern on the front at time~$t$+1. Remark that
the {\tt B}-cell of the pattern {\tt BW} on ${\cal F}_{N_{t+1}}$ is isolated. The same
arguments can be repeated to the black son of~$\sigma$+1. Consequently, a pattern 
{\tt BW} where the {\tt B}-cell is isolated on the front at time~$t$ generates a 
sequence of such patterns on each front 
at time~$t$+$k$, with $k$ being a positive integer, the {\tt B}-cell of such a pattern
being isolated and being the black son of the {\tt W}-cell of the same pattern on the 
previous front.
We can call this sequence a {\bf line of patterns {\tt BW}}. Accordingly, if there
are $k$ patterns {\tt BW} on the front at time~$t$, each of them generates a line
of patterns {\tt BW} on the successive fronts after time~$t$.

From now on, consider the case when the program of~$A$ contains the rule \ab. 

If a {\tt B}-cell occurs on the front at time~$t$ on the node~$\nu$, the white sons of~$\nu$
become {\tt B}-cells at time~$t$+1, as seen in the proof of Theorem~\ref{decid}. 
Accordingly, not only the front at time~$t$+1 contains a {\tt B}-cell, it also contains 
the pattern {\tt BB}. If $\nu$+1 is a {\tt W}-cell, its black son is a {\tt B}-cell if 
and only if the program of~$A$ contains the rule \wb. In that case the front at time~$t$+1 
contains the pattern {\tt BBB}. The occurrence of the pattern {\tt BB} on the front raises 
the question of which rule \bb{} or \bbw{} belongs to the program of~$A$.

The easiest case to analyze is the case when together with the rule \ab{} we also have
the rules \wb{} and \bb.
\vskip 10pt
\vtop{
\ligne{\hfill
\includegraphics[scale=0.3]{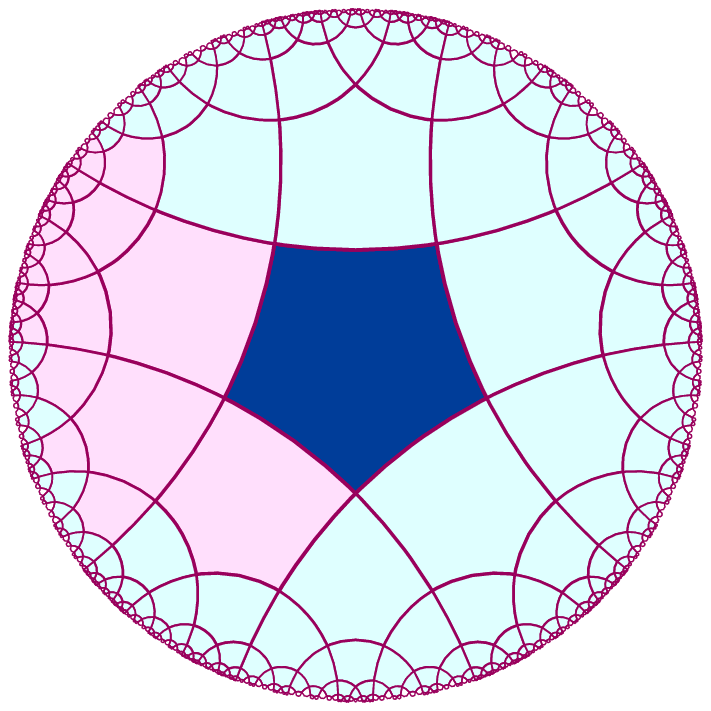}\hfill
\includegraphics[scale=0.3]{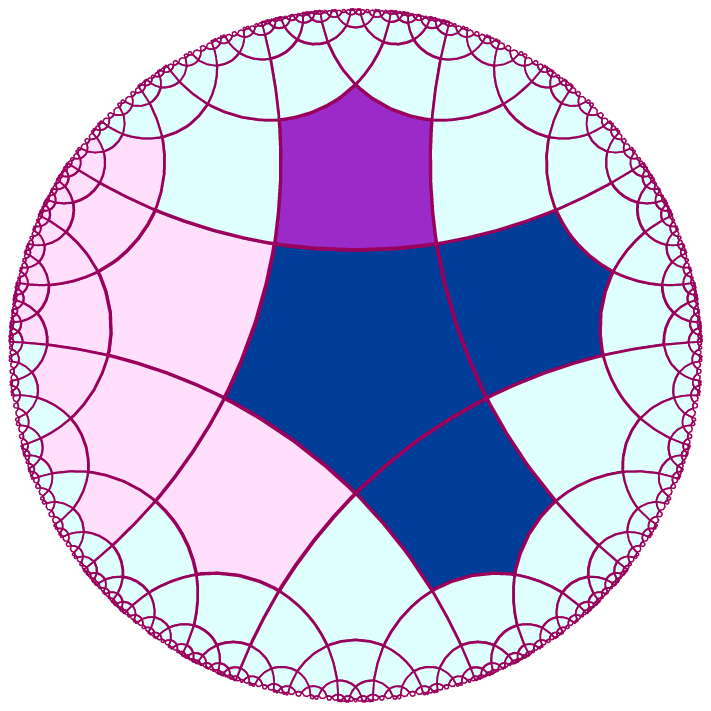}\hfill
\includegraphics[scale=0.3]{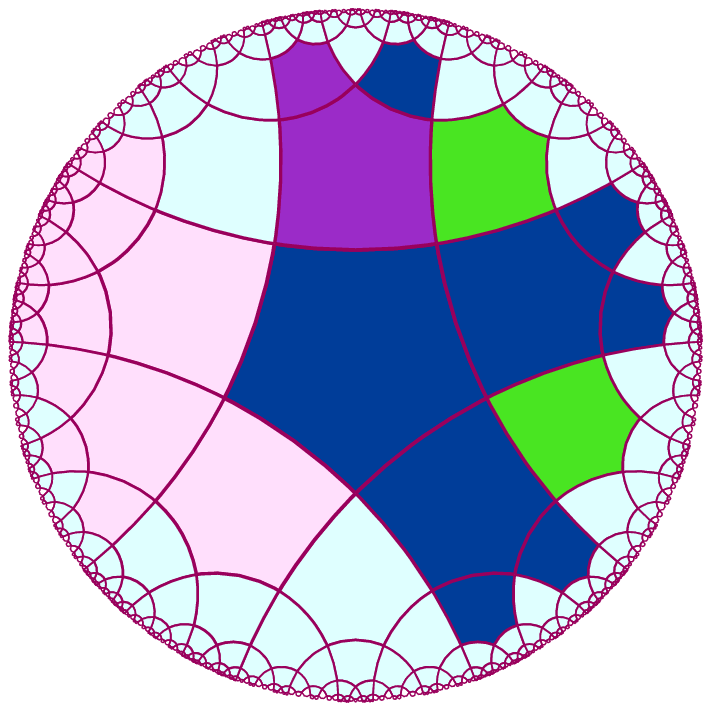}\hfill
\includegraphics[scale=0.3]{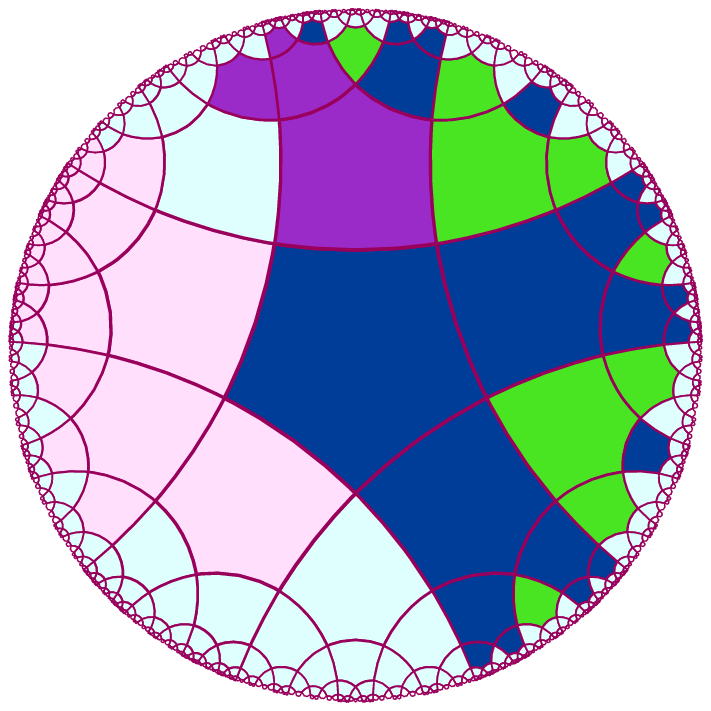}\hfill
\includegraphics[scale=0.3]{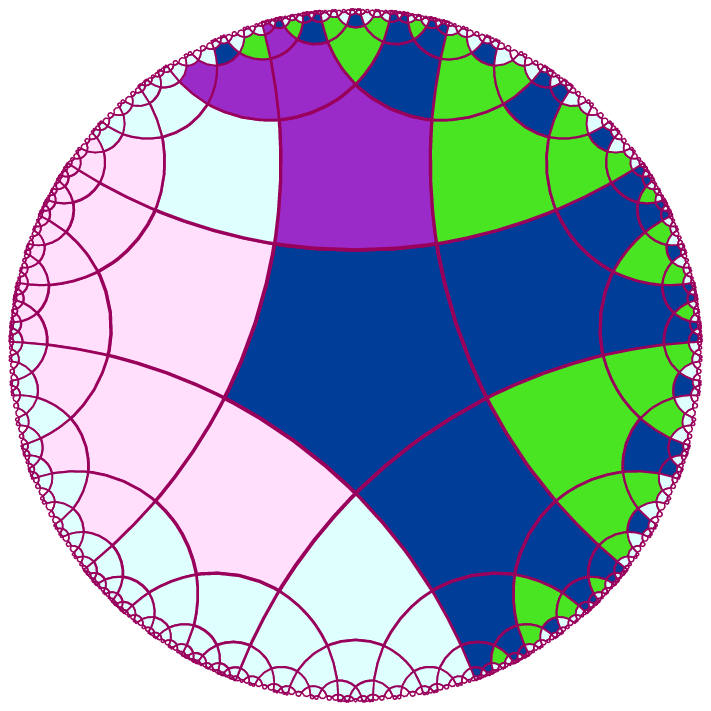}\hfill
}
\begin{fig}\label{FabwbbbF}
\leurre
The program contains the rules \ab, \wb{} and \bb.
From left to right, times $0$, $1$, $2$, $3$ and $4$. It is assumed that once a node
is a {\tt B}-cell, it remains in this situation. The light pink cells represent
the circles which are behind the front at time~$t$.
\end{fig}
}

In the pictures of Figure~\ref{FabwbbbF}, the result of applying the rules \ab, \wb{}
and \bb, respectively, yields the cells in blue, purple and green, respectively.
Clearly, the white neighbour of the purple neighbour of the central cell is its father,
see the pictures for times~1, 2, 3 and~4. It is assumed that the state {\tt B} is
permanent: once a cell gets that state, it remains unchanged. The figure also assumes that
we start from a single {\tt B}-cell on the front at time~0. That cell is placed as the
central cell of the Figures~\ref{FabwbbbF}, \ref{FabwbbbL} and~\ref{FabwbbbwL} in order
to focus the attention on the evolution of the computation from that cell.

   In Figure~\ref{FabwbbbL}, contrarily to Figure~\ref{FabwbbbF}, it is assumed that
a {\tt B}-cell at time~$t$ becomes quiescent and remains in that state later on.
Note that this representation allows us to better see the propagation of the front
in the case of the motions ruled by the occurrence of the rule \ab{} in the program of~$A$.
As we assume that the rules \wb{} and \bb{} also belong to the program of~$A$,
we can easily see that the sons of a {\tt B}-cell in node~$\nu$ at time~$t$ are 
{\tt B}-cells at time~$t$+1 whatever the states at time~$t$ of the nodes $\nu$$-$1
and $\nu$+1. Accordingly, on the front at time~$t$+$k$ the {\tt B}-cells occupy
at least the whole level~$k$ of the Fibonacci tree rooted at $\nu$, whether $\nu$ is
a black node or a white one.

\vskip 10pt
\vtop{
\ligne{\hfill
\includegraphics[scale=0.3]{propa_front1_0.ps}\hfill
\includegraphics[scale=0.3]{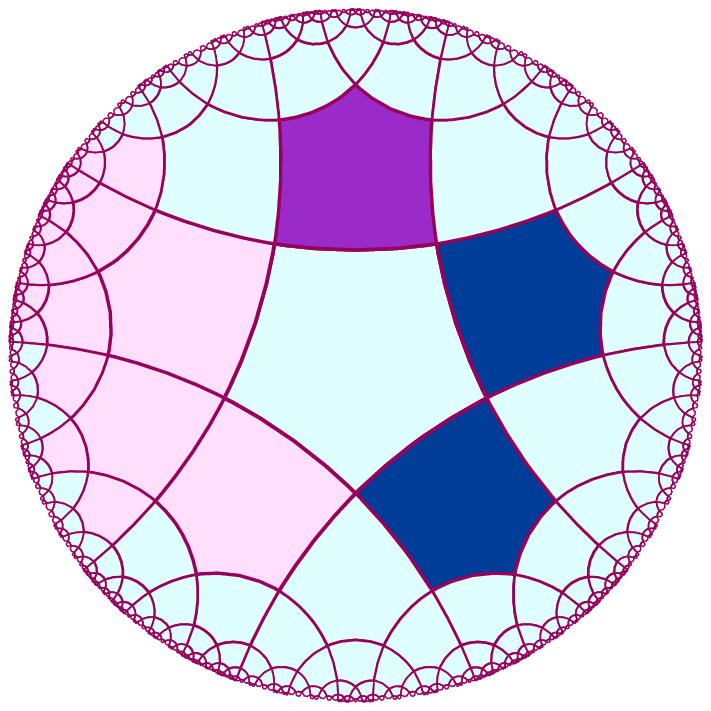}\hfill
\includegraphics[scale=0.3]{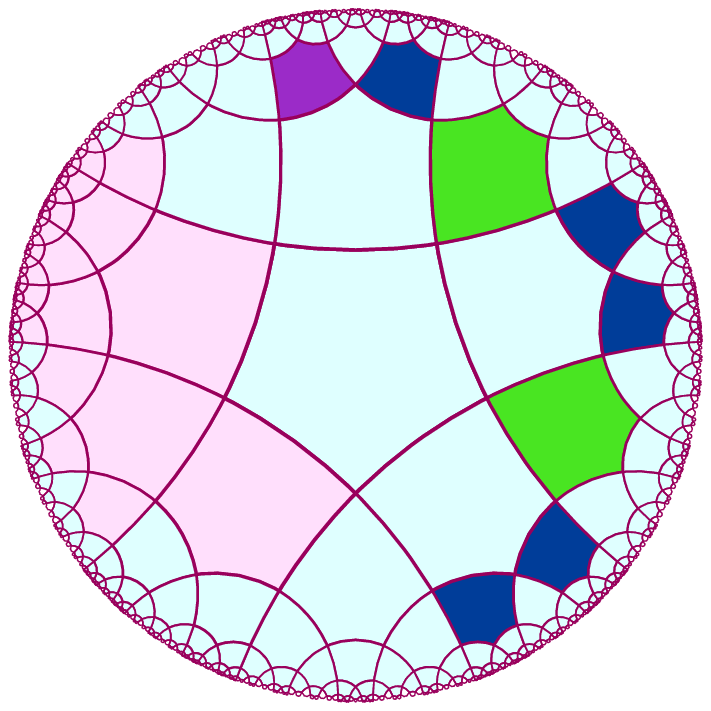}\hfill
\includegraphics[scale=0.3]{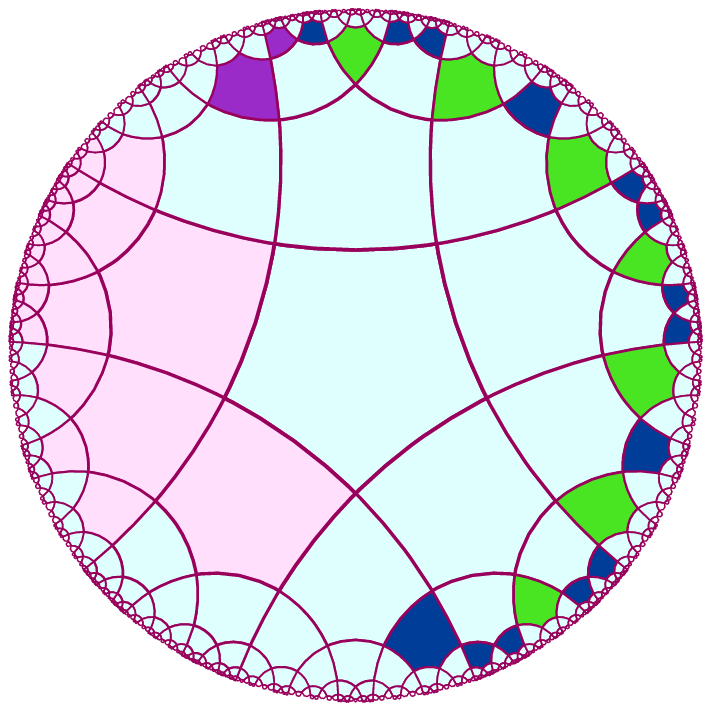}\hfill
\includegraphics[scale=0.3]{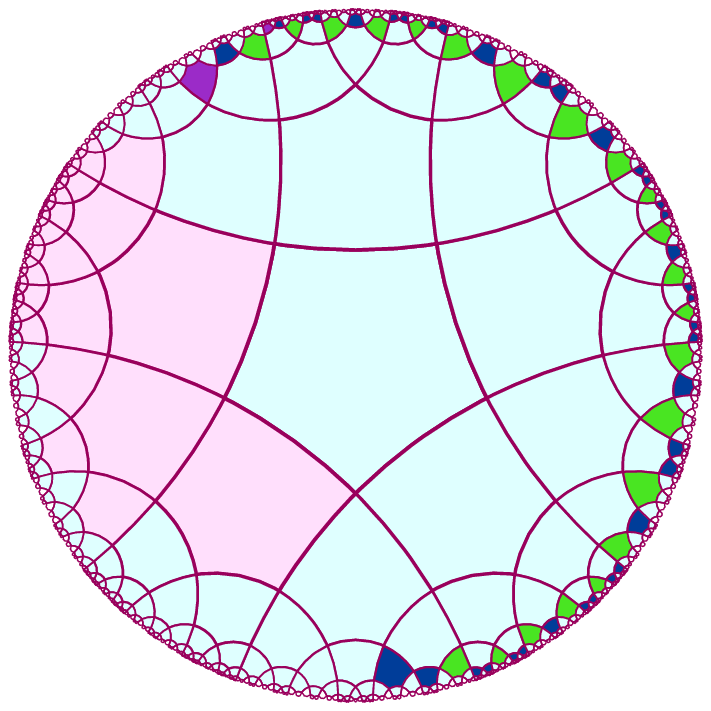}\hfill
}
\begin{fig}\label{FabwbbbL}
\leurre
The program contains the rules \ab, \wb{} and \bb.
From left to right, times $0$, $1$, $2$, $3$ and $4$. It is assumed that when a node
is a {\tt B}-cell, it becomes a quiescent cell at the next time.
\end{fig}
}
\vskip 5pt
   Still assume that we have the rules \ab{} and \wb, but that we have the rule \bbw.
\vskip 10pt
\vtop{
\ligne{\hfill
\includegraphics[scale=0.3]{propa_front1_0.ps}\hfill
\includegraphics[scale=0.3]{propa_front2_1.ps}\hfill
\includegraphics[scale=0.3]{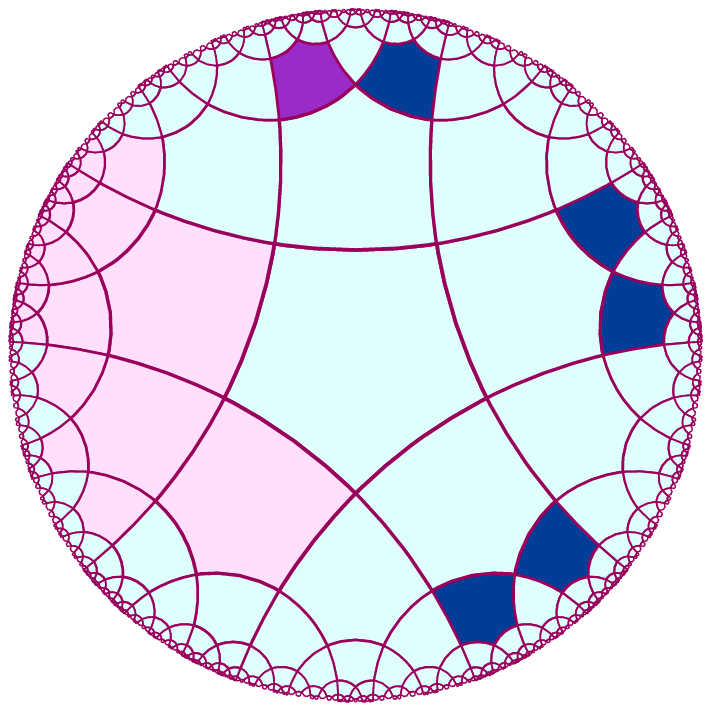}\hfill
\includegraphics[scale=0.3]{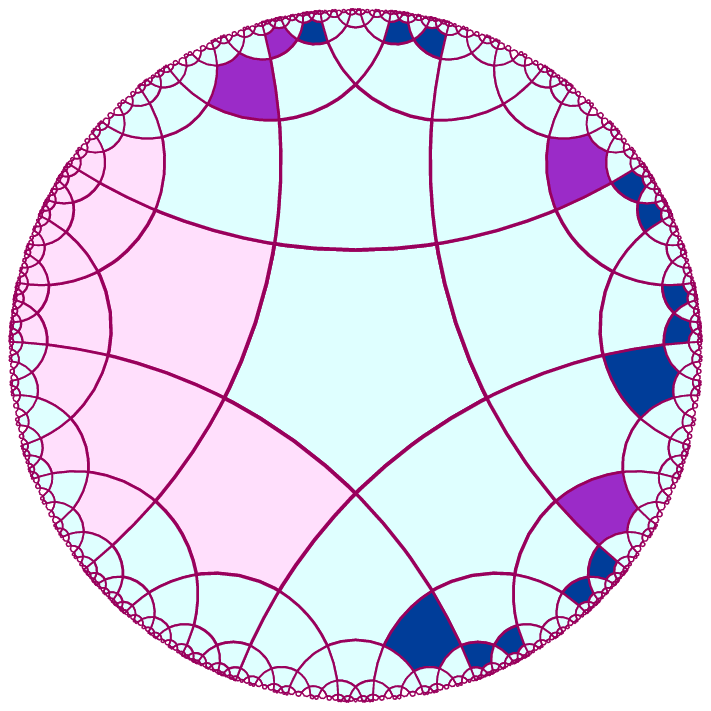}\hfill
\includegraphics[scale=0.3]{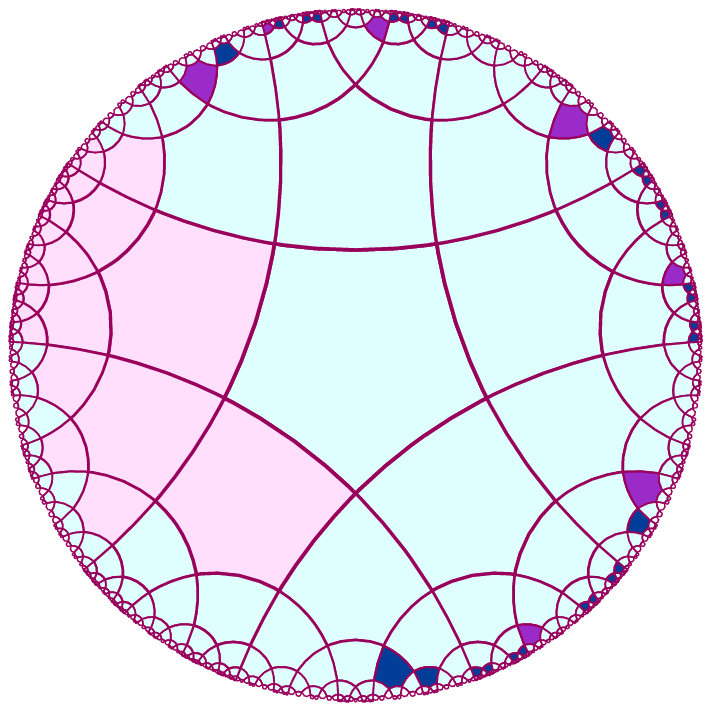}\hfill
}
\begin{fig}\label{FabwbbbwL}
\leurre
The program contains the rules \ab, \wb{} and \bbw.
From left to right, times $0$, $1$, $2$, $3$ and $4$. It is assumed that when a node
is a {\tt B}-cell, it becomes a quiescent cell at the next time.
\end{fig}
}
\vskip 5pt
   Figure~\ref{FabwbbbwL} illustrates the propagation of the front in that case, starting
with a single {\tt B}-cell on the front at the initial time. The picture at time~3 in
that figure indicates that the pattern {\tt BBB} appears on the front of that time and
the picture at time~4 seems to indicate the same property and that no new pattern 
appears. Let us prove this property.

\begin{lem}\label{whites}
Let $A$ be a deterministic cellular automaton on the pentagrid with two states, one of
them being quiescent. Assume that the program of~$A$ contains the rule~\ab.
The states of the cells attached to the white sons of a white node 
$\nu$ at time~$t$$+$$1$ are the state of the cell of~$\nu$ at time~$t$.
\end{lem}

\noindent
Proof of the lemma. Indeed, let $\nu$ be a cell of the front at time~$t$ supported
by a white node. Its white sons belong to ${\cal F}_{N_t+1}$ and, as white nodes,
they have one neighbour on ${\cal F}_{N_t}$ and four of them on ${\cal F}_{N_t+2}$.
At time~$t$, those four neighbours are {\tt W}-cells by definition of the front at
time~$t$, so that the quiescent rule applies if $\nu$ is a {\tt W}-cell and the rule~\ab{}
applies if $\nu$ is a {\tt B}-cell. In both cases, we get the conclusion of the lemma.
\hfill$\Box$

This lemma shows that among the sons of a white node on the front, two of them always 
have the same state at the next time. We can now state:

\begin{lem}\label{nofour}
Let $A$ be a deterministic cellular automaton on the pentagrid with two states, one of
them being quiescent. Assume that the program of~$A$ contains the rules~\ab, \wb{} 
and~\bbw. Then, the front at time~$t$, with $t\geq3$ does not contain neither the pattern
{\tt WBW} nor the pattern {\tt BBBB}.
\end{lem}

\noindent
Proof of the lemma. Assume that the pattern {\tt WBW} occurs at time~$t$+2. 
Let $\nu$, $\nu$+1 and $\nu$+2 be the nodes supporting that pattern. From the rules and from
Lemma~\ref{whites}, the nodes $\nu$, $\nu$+1 and~$\nu$+2 cannot
have the same father which should be a white node. Accordingly, the fathers of~$\nu$+1 
and~$\nu$+2 are different, say $\varphi$+1 and $\varphi$+2. Assume that $\varphi$+2 is 
the father of~$\nu$+1 and $\nu$+2, so that $\nu$+1 is a black node and $\nu$+2 is a 
white one. Also, $\nu$ must be a white son of~$\varphi$+1. By Lemma~\ref{whites}, 
both $\varphi$+1 and $\varphi$+2 should be {\tt W}-cells, so that by the quiescent rule,
$\nu$+1 should be a {\tt W}-cell, a contradiction with our assumption.
And so, we have that $\varphi$+1 is the father of 
$\nu$ and $\nu$+1 and that $\varphi$+2 is the father of $\nu$+2. 

Then, $\varphi$+1, which we consider on the front at time~$t$+1 cannot be a white node
as its white sons would bear different states at time~$t$+2. So $\varphi$+1 is a black node
and $\nu$ is its black son and $\nu$+1 is its white one. Accordingly, $\varphi$+1 is 
a {\tt B}-cell at time~$t$+1.
As $\nu$+2 is a {\tt W}-cell at time~$t$+2, $\varphi$+2 must also be a {\tt B}-cell
at time~$t$+1. 
Let us look at what happens at time~$t$+3.
Let $\sigma$ be the white son of~$\nu$ which is a black node. Then, 
the rule~\ab{} apply to~$\sigma$, $\sigma$+1, $\sigma$+2
and $\sigma$+3 producing the pattern {\tt WBBB}.
Now, the rule~\wb{} applies to~$\sigma$+4 as that node is a black one, and by 
Lemma~\ref{whites}, $\sigma$+5 is quiescent, so that starting from~$\sigma$,
the sons of~$\nu$, $\nu$+1 and~$\nu$+2 produce the state pattern
{\tt WBBBBW} at time~$t$+3.
Applying the rules in a similar way, at time~$t$+4, starting from
the rightmost son of~$\sigma$, we obtain the pattern {\tt WBBWBBWBBWBBW} where the rightmost
{\tt W} is the first white son of~$\sigma$+5.

Now, consider the case of a pattern {\tt WBBBBW} on the front at time~$t$+1, and let
$\nu$ be the node which gives the leftmost {\tt W} at that time. By Lemma~\ref{whites},
$\nu$, $\nu$+1 and $\nu$+2 cannot be the sons of a white node~$\varphi$. 
A similar contradiction would 
occur if we assume that $\nu$ and $\nu$+1 are sons of a black node. We conclude that 
$\nu$ is the rightmost son of a node~$\varphi$. If we assume that $\nu$+2 and $\nu$+3 
are the sons of $\varphi$+1 which should accordingly be a black node supporting 
a {\tt B}-cell, we have a contradiction between the state of $\varphi$+2 at time~$t$, 
which would be a white node and 
that of~$\nu$+3 and $\nu$+4 at time~$t$+1. Accordingly, $\varphi$+1 must be a white node 
and $\nu$+4 and $\nu$+5 are sons of $\varphi$+2, so that we find the situation associated 
with the pattern {\tt WBW}.

We have seen that the pattern produced by the sons of the nodes supporting
{\tt WBBBBW} does not contain neither {\tt WBW}
nor {\tt BBBB}. It contains four occurrences of {\tt BB}, two of them being separated
by a single {\tt W}.

   Now, let us look at the pattern {\tt WBBW} which we assume to be on the front at 
time~$t$+1. Let $\nu$ be the node which supports the
left-hand side {\tt W} of the pattern. The nodes~$\nu$, $\nu$+1 and $\nu$+2 can be the
sons of a white node~$\varphi$+1 which is necessarily a {\tt B}-cell at time~$t$. 
As $\nu$ is a {\tt W}-cell at time~$t$+1, $\varphi$ must be a {\tt B}-cell at time~$t$.
This indicates which kind of nodes are~$\nu$, $\nu$+1 and $\nu$+2 and, clearly, 
$\nu$+3 is a black node. Applying the rules to the sons of these nodes,
we get that, from $\sigma$ is the rightmost son of~$\nu$ until the leftmost white son
of~$\nu$+3, the nodes $\nu$+$i$ produce the pattern
{\tt WBBBWBBBW} at time~$t$+2.

   However, $\nu$+1, $\nu$+2 and $\nu$+3 cannot be the sons of a white node as  
$\nu$+2 and $\nu$+3 have different states. Another disposition for the fathers of
the node we have seen is that the father of~$\nu$, say $\varphi$, is a black node, so
that $\varphi$+1 is a white one. Necessarily, $\varphi$ is a {\tt B}-cell at time~$t$
and $\varphi$+1 is a white one. Looking at the sons of the nodes $\nu$, $\nu$+1,
$\nu$+2 and~$\nu$+3, starting from the rightmost son~$\sigma$ of~$\nu$ until the
leftmost white son of~$\nu$+3 we find this time the pattern {\tt WBBBWBBW}.

   Let us now consider the pattern {\tt WBBBW} on the front at time~$t$+2. 

Again, let $\nu$ be the node which supports the leftmost {\tt W}-cell of this pattern. We 
can see that the nodes $\nu$, $\nu$+1 and $\nu$+2 can be the sons of a node~$\varphi$ which
must be a {\tt B}-cell at time~$t$ while the node~$\varphi$+1 must be a {\tt W}-cell
at the same time. It is not difficult to see that under that assumption on $\varphi$ and
$\varphi$+1 with respect to the nodes $\nu$+$i$, if $\sigma$ is the
rightmost son of~$\nu$, we get the pattern {\tt WBBBWBBWBBW} on the front at time~$t$+2
until the leftmost white son of $\nu$+4.

   If $\nu$ and $\nu$+1 would be the sons of a black node~$\varphi$ while the other 
nodes $\nu$+$i$ would be the sons of the necessarily white node~$\varphi$+1. 
The nodes~$\nu$+3 and~$\nu$+4 have different colours at time~$t$+1, a contradiction
with Lemma~\ref{whites}. And so, another configuration which this time is possible, 
is that the nodes $\nu$+$i$ we consider have three fathers: $\varphi$ is the father 
of~$\nu$ only, $\varphi$+1 is a white node or a black one, respectively, it does not 
matter, and $\varphi$+2 is the father of $\nu$+4, or of~$nu$+3 and~$\nu$+4, respectively.
In both cases, $\nu$+1 is a black node, it is the important point. It can be
checked that in both cases, if $\sigma$ is the rightmost son of~$\nu$, the pattern
on the front at time~$t$+2 starting from $\sigma$ and ending on the first white son
of~$\nu$+4 is {\tt WBBWBBWBBBW}.

   With this analysis, the proof of Lemma~\ref{nofour} is completed.\hfill$\Box$

We have analyzed the situation when the program of~$A$ contains the rules~\ab{}
and~\wb. With programs containing the rule~\ab, we remain to consider the case
of the rule~\wbw. Figure~\ref{FabwbwL} illustrates the propagation of the front
whatever the rule~\bb{} or \bbw.
\vskip 10pt
\vtop{
\ligne{\hfill
\includegraphics[scale=0.3]{propa_front1_0.ps}\hfill
\includegraphics[scale=0.3]{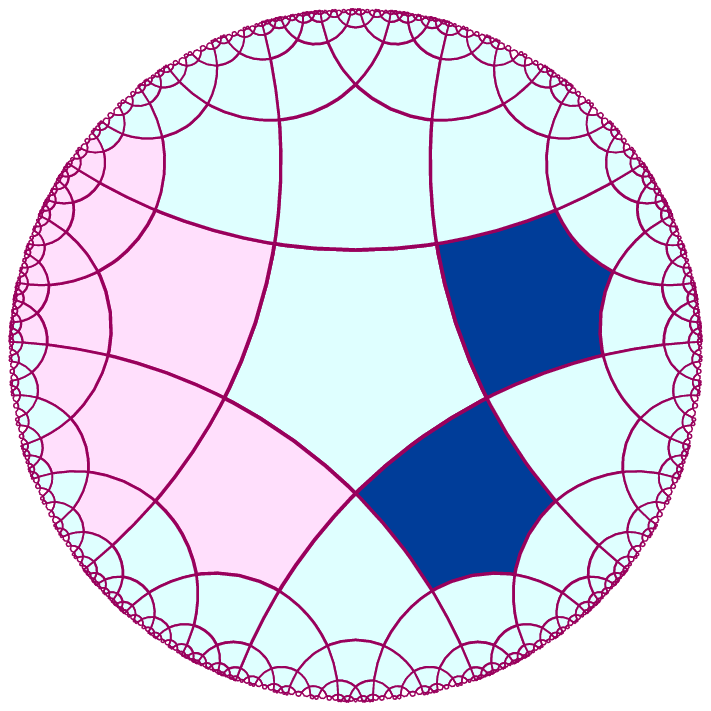}\hfill
\includegraphics[scale=0.3]{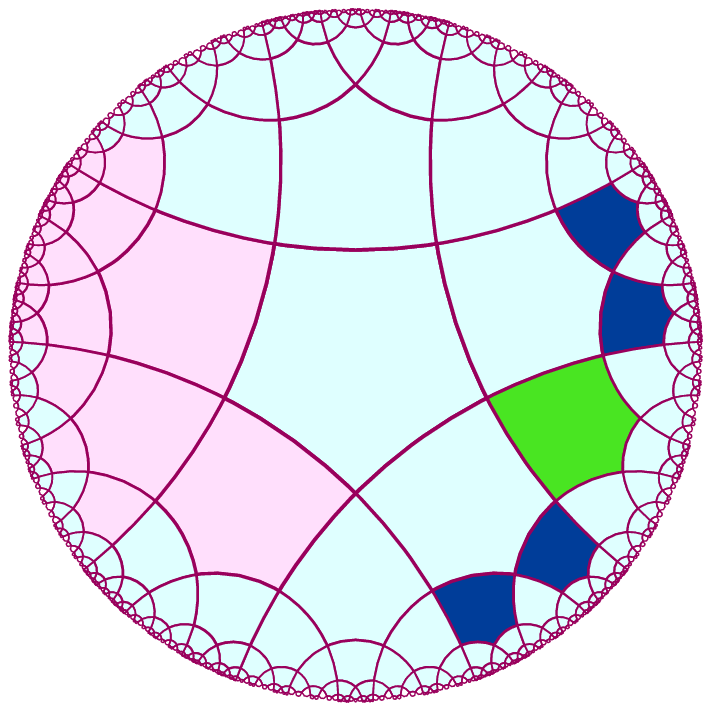}\hfill
\includegraphics[scale=0.3]{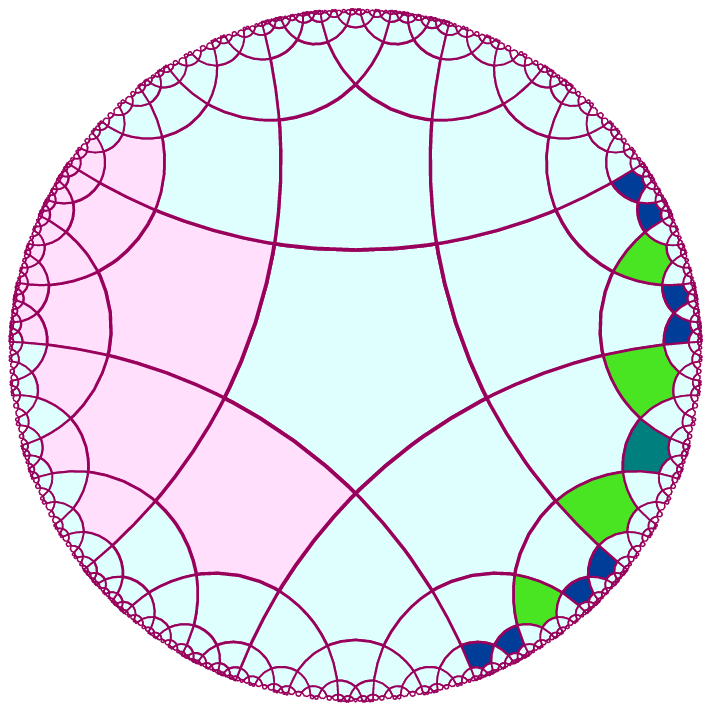}\hfill
\includegraphics[scale=0.3]{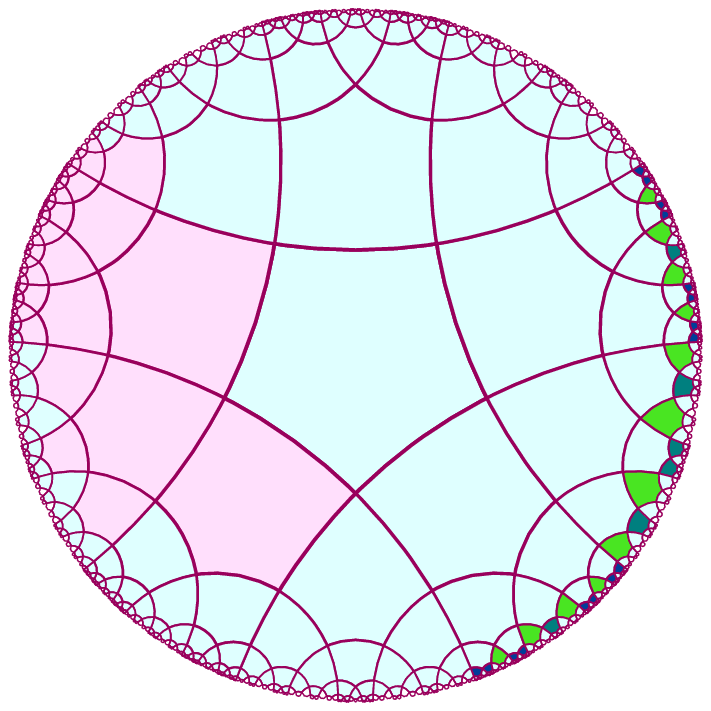}\hfill
}
\begin{fig}\label{FabwbwL}
\leurre
The program contains the rules \ab{} and \wbw. In green and lighter dark blue the
{\tt B}-cells produced when using the rule~\bb{} too. When the program contains the
rule~\bbw, the {\tt B}-cells are restricted to the dark blue cells.
From left to right, times $0$, $1$, $2$, $3$ and $4$. It is assumed that when a node
is a {\tt B}-cell, it becomes a quiescent cell at the next time.
\end{fig}
}

In fact, the figure illustrates both cases: as mentioned in the caption of the figure,
a different coloration is applied to the cells produced directly by the application
of the rule~\bb{} or to further applications of all rules in the tree rooted at the node
where a first application of the rule~\bb{} was performed. When the rules \ab, \wbw{} and
\bb{} are applied starting from an isolated {\tt B}-cell supported by a node~$\nu$
on the front at time~$t$, the evolution of the computation concerns the Fibonacci tree
rooted at~$\nu$ and on the front at time~$t$+$k$, the trace of that computation is the
whole level~$k$ of that tree. Say that a node~$\nu$ is {\bf hereditary white} if there
is a sequence of~$k$ white nodes~$\nu_i$, with \hbox{$i\in\{1..k\}$} such that
$\nu_{i+1}$ is a white son of~$\nu_i$ with \hbox{$i\in\{1..k$$-$$1\}$} and $\nu=\nu_k$.
When the rule~\bbw{} is used in place of the rule~\bb,
the trace is restricted to hereditary white nodes only.

We can summarize our analysis by appending the Table~\ref{Tpropa} to Table~\ref{Tdecid}.
The table assumes that we start from a {\tt B}-cell supported by a white node of the
front at time~$t$. In order to better analyze the patterns, we remind the reader
that the number of nodes on the level~$k$ of a Fibonacci tree rooted at a white node,
a black node, respectively, is $f_{2k+1}$, $f_2k$, respectively.

\ligne{\hfill
\vtop{\leftskip 0pt\parindent 0pt\hsize=320pt
\begin{tab}\label{Tpropa}
\leurre
Patterns on the front at time~$t$$+$$k$ when the program of~$A$ contains
the rule~\ab{} starting from a {\tt B}-cell in a white node of the front at time~$t$. 
\end{tab}
\vskip -8pt
\LLlarge=70pt
\ligne{\hfill
\vtop{\leftskip 0pt\parindent 0pt\hsize=260pt
\ligne{\hfill
\hbox to \LLlarge {\hfill rules\hfill} \hbox to 160pt{\hfill patterns at $t$+$k$\hfill}
\hfill}
\ligne{\hfill
\hbox to \LLlarge {\hfill \ab, \wb, \bb\hfill} 
\hbox to 160pt{\hfill\tt WB$^{f_{2k+1}+f_{2k-2}} $W\hfill}
\hfill}
\ligne{\hfill
\hbox to \LLlarge {\hfill \ab, \wb, \bbw\hfill} 
\hbox to 160pt{\hfill{\tt WBBW}, {\tt WBBBW} in a range\hfill}
\hfill}
\ligne{\hfill
\hbox to \LLlarge {\hfill} \hbox to 160pt{\hfill wider than $f_{2k+1}+f_{2k-2}$ nodes\hfill}
\hfill}
\ligne{\hfill
\hbox to \LLlarge {\hfill \ab, \wbw, \bb\hfill} 
\hbox to 160pt{\hfill\tt WB$^{f_{2k+1}} $W\hfill}
\hfill}
\ligne{\hfill
\hbox to \LLlarge {\hfill \ab, \wbw, \bbw\hfill} 
\hbox to 160pt{\hfill {\tt B} on hereditary white nodes\hfill}
\hfill}
\ligne{\hfill
\hbox to \LLlarge {\hfill} 
\hbox to 160pt{\hfill in a range wider than $f_{2k+1}$ nodes\hfill}
\hfill}
}
\hfill}
}
\hfill}
\vskip 10pt
\vtop{
\ligne{\hfill
\includegraphics[scale=0.3]{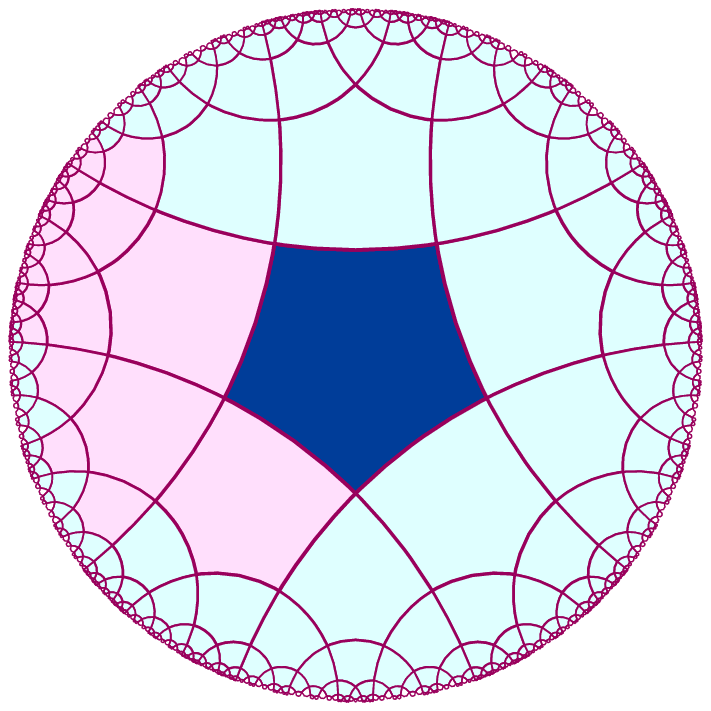}\hfill
\includegraphics[scale=0.3]{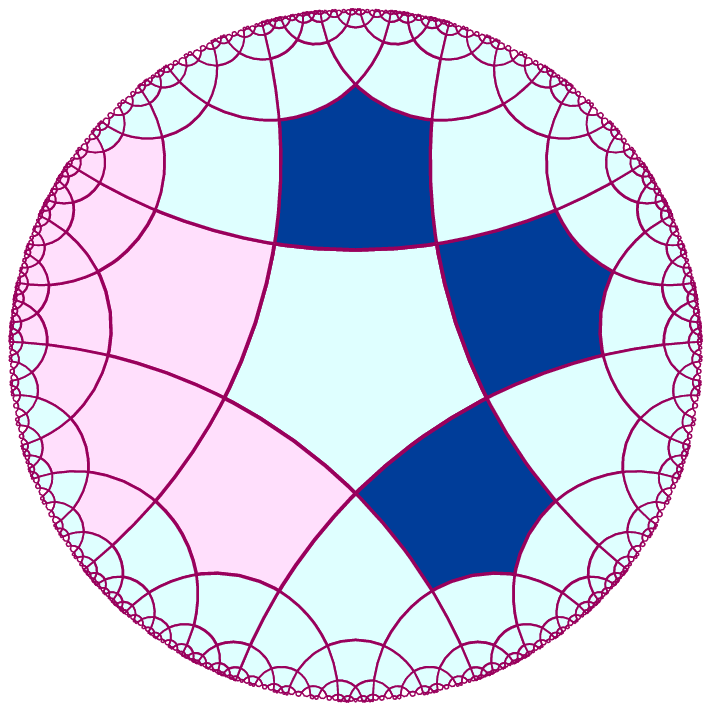}\hfill
\includegraphics[scale=0.3]{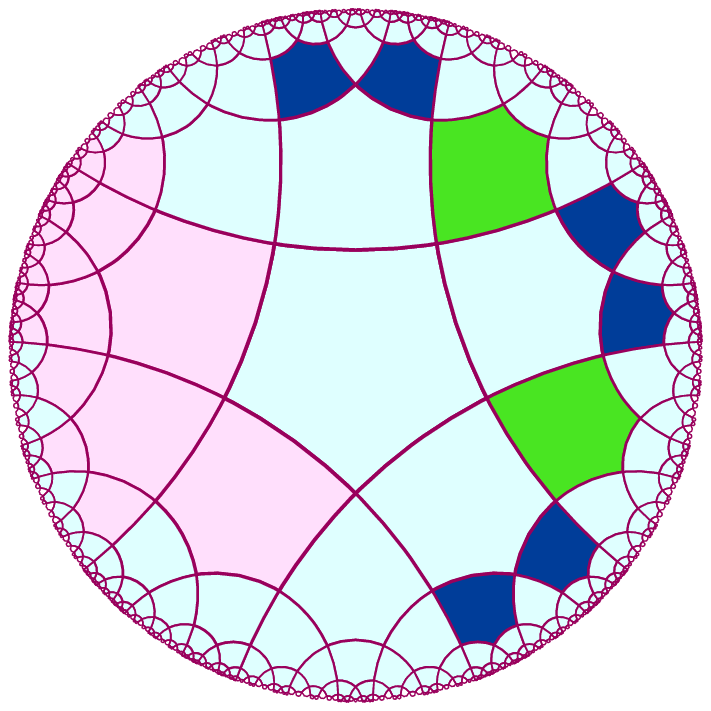}\hfill
\includegraphics[scale=0.3]{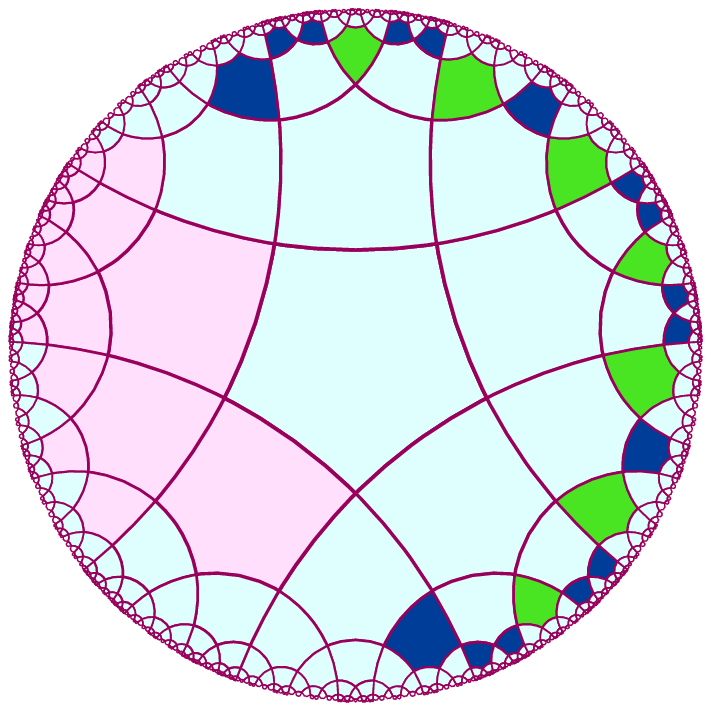}\hfill
\includegraphics[scale=0.3]{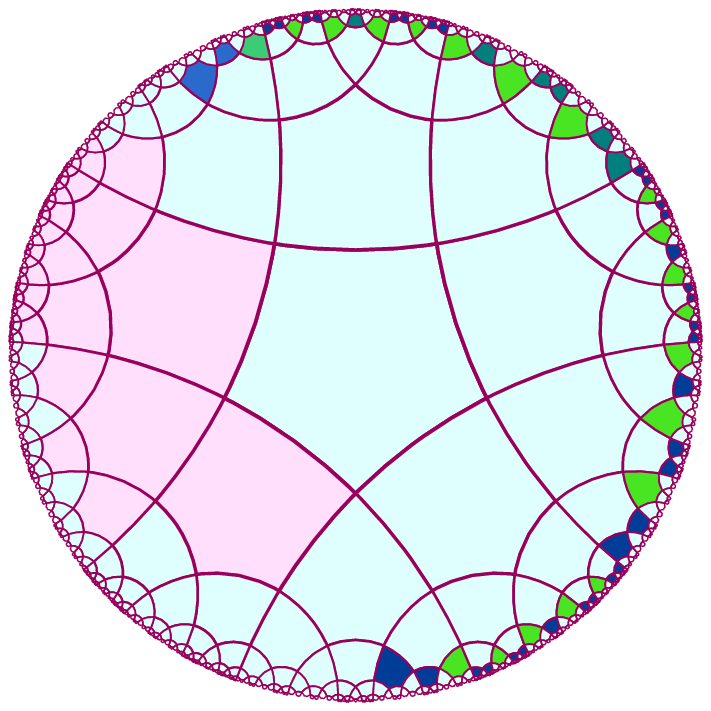}\hfill
}
\begin{fig}\label{FriabbbL}
\leurre
The program contains the rotation invariant rules \ab{} and \bb. In green the
{\tt B}-cells produced when using the rule~\bb{} too. 
From left to right, times $0$, $1$, $2$, $3$ and $4$. It is assumed that when a node
is a {\tt B}-cell, it becomes a quiescent cell at the next time.
\end{fig}
}
\vskip 10pt
We remain to append an information regarding the case when the program of~$A$ is rotation
invariant. The first remark is that in such a situation, there is no difference between
the rules \ab{} and~\wb{} as well as between the rules \abw{} and~\wbw. As we assume the
rule~\ab, there is no consideration of a rule~\wbw. This also means that in a situation
where we applied the rule~\wb{} when rotation invariance is relaxed, in the case of rotation
invariance we apply the rule~\ab. However, note that the rules \ab{} and \wbw{} are 
contradictory under rotation invariance as in that case, \wbw{} is the same rule as~\abw{}
which, by definition, is opposite to~\ab. And so, we are concerned with
the first two rows of Table~\ref{Tpropa}. However, there is a special phenomenon which
occurs here and may not occur in the situation where we deal when the rotation invariance
does not take place. It is illustrated by Figures~\ref{FriabbbL} and~\ref{FriabbbwL}.

In Figure~\ref{FriabbbL}, we assume that besides the rule~\ab, the rule~\bb{} too belongs
to the program of~$A$. In this case too, the rules~\ab{} and~\wb{} are the same up to
a circular permutation on the neighbours.

Comparing Figure~\ref{FabwbbbL} with Figure~\ref{FriabbbL} on one hand and 
Figure~\ref{FabwbbbwL} with Figure~\ref{FriabbbwL} on the other hand  we can see in both
cases that the {\tt B}-cells are at the same places during the propagation.
\vskip 10pt
\vtop{
\ligne{\hfill
\includegraphics[scale=0.3]{propa_frontri_0.ps}\hfill
\includegraphics[scale=0.3]{propa_frontri_1.ps}\hfill
\includegraphics[scale=0.3]{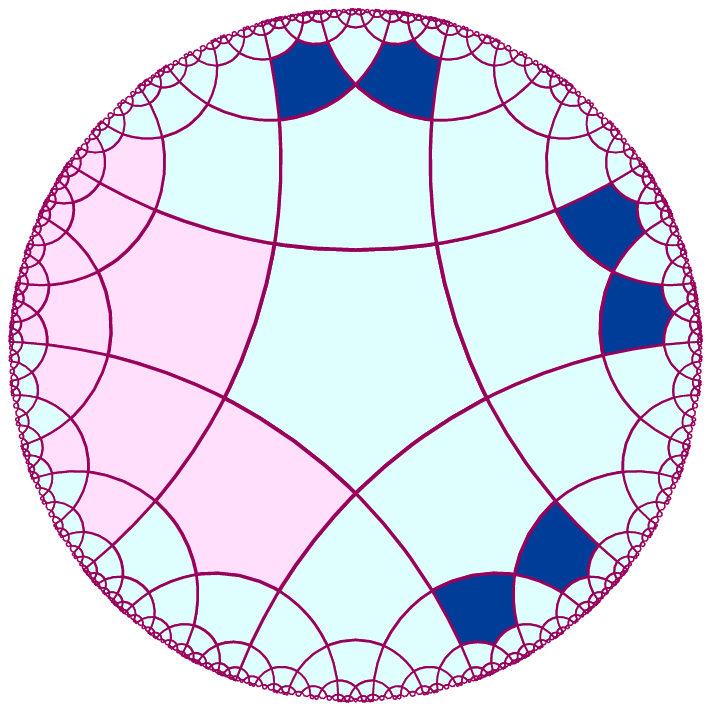}\hfill
\includegraphics[scale=0.3]{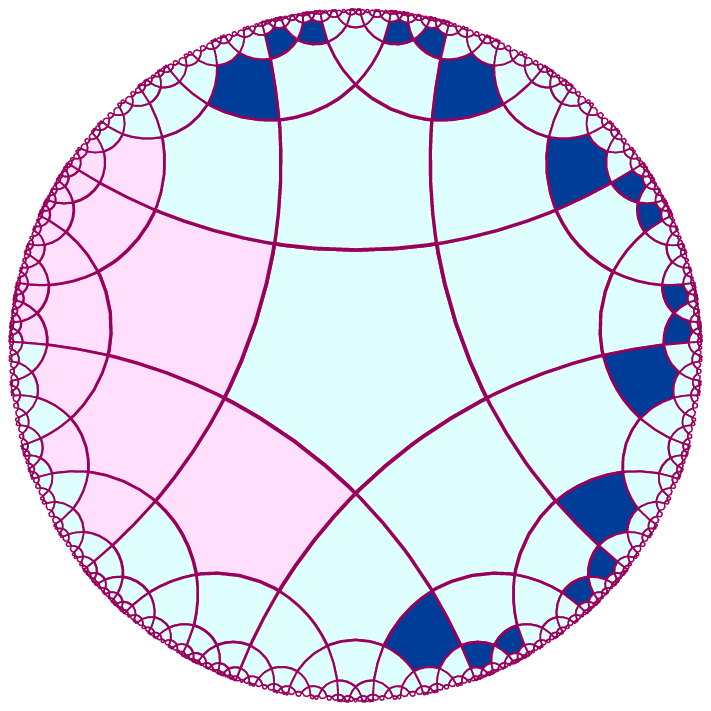}\hfill
\includegraphics[scale=0.3]{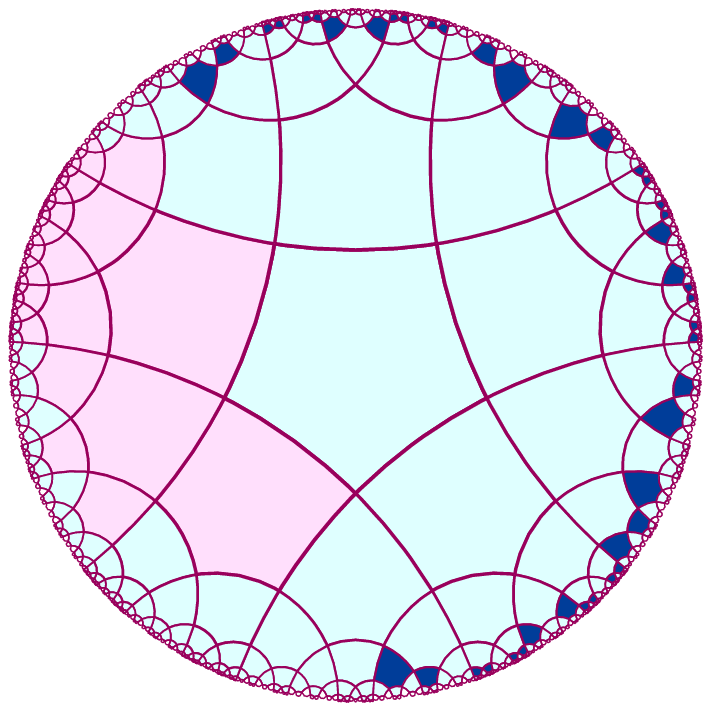}\hfill
}
\begin{fig}\label{FriabbbwL}
\leurre
The program contains the rotation invariant rules \ab{} and \bbw. 
From left to right, times $0$, $1$, $2$, $3$ and $4$. It is assumed that when a node
is a {\tt B}-cell, it becomes a quiescent cell at the next time.
\end{fig}
}

\section{Conclusion}
\label{conclude}

   Of course, the first question is what can be said for three states? That issue is
more difficult. We already have seen a rather difficult situation in the proof
of Theorem~\ref{decid} when it could happen that the front enters a blinking between
all cells in the state~{\tt B} and all of them in the state~{\tt W} at the next time
and conversely. As then the front remains at the same place during a certain time, the 
discussion was how long such a blinking might last. A worse situation occurs with
three states for something which we could ignore with two states: the point is what happens
behind the front? In fact, in case a node changes its state from~{\tt W} to~{\tt B} behind
the front, the worse thing it might happen is that another line of {\tt B}-cells might
propagate but, in that case, another line already occurred so that we are in the situation
of a constant advance of the front. Accordingly, it does not change the situation for what
is the halting of the computation. 

The things are different with three states. Let the three states being {\tt W}, {\tt B}
and~{\tt R}, where {\tt W} is the quiescent state which is associated to the quiescent
rule possessed by the program of our cellular automaton. As a third
state enters the play, we may have the following rules:
\def\bwb{{\hbox{\footnotesize\tt BWB}}}
\def\bwr{{\hbox{\footnotesize\tt BWR}}}
\def\rwb{{\hbox{\footnotesize\tt RWB}}}
\def\rwr{{\hbox{\footnotesize\tt RWR}}}
\def\rww{{\hbox{\footnotesize\tt RWW}}}
\vskip 5pt
\ligne{\hfill
\bwb:\hskip 5pt 
\hbox{\tt$\underline{\hbox{\tt W}}$BW$^4\underline{\hbox{\tt B}}$}, 
\hskip 10pt\bwr:\hskip 5pt 
\hbox{\tt$\underline{\hbox{\tt W}}$BW$^4\underline{\hbox{\tt R}}$}, 
\hskip 10pt\rwb:\hskip 5pt 
\hbox{\tt$\underline{\hbox{\tt W}}$RW$^4\underline{\hbox{\tt B}}$}.
\hskip 10pt\rwr:\hskip 5pt 
\hbox{\tt$\underline{\hbox{\tt W}}$RW$^4\underline{\hbox{\tt R}}$}.
\hskip 10pt\rww:\hskip 5pt 
\hbox{\tt$\underline{\hbox{\tt W}}$RW$^4\underline{\hbox{\tt W}}$}.
\hfill}
\vskip 5pt
Clearly, if we have the rule~\bwb, or the rule~\rwr, we have a constant progression
of the front once a non-quiescent cell occurs on the front. A similar conclusion
occurs if we have both rules~\bwr{} and~\rwb: they call each other in some sense,
again once a non-quiescent cell occurs on the front. What happens if, instead of both
rules~\bwr{} and~\rwb{} we have, for instance, both rules~\bwr{} and~\rww?
In that case, assume that the rule~\bwr{} applies to the node~$\varphi$ of the front at 
time~$t$. Let~$\nu$ be the first white son of~$\varphi$ and let~$\sigma$ be the first 
white son of~$\nu$. 
Then, at time~$t$+1 $\nu$ becomes an {\tt R}-cell and, due to~\rww, at time~$t$+2, 
$\sigma$ becomes
remains a quiescent cell. Now, it may happen that at time~$t$+2, $\nu$ becomes
a {\tt B}-cell. In that case, $\sigma$ becomes an {\tt R}-cell at time~$t$+3. However,
even if the transformation of $\nu$ from an \hbox{{\tt R}-cell} to a {\tt B}-cell happens at 
time~$t$+2, we are not guaranteed that the same transformation will happen for 
$\sigma$ at time~$t$+4. The reason is that in those cases, the transformation depends
on what happened behind the front. Note that in our discussions with a single non-quiescent
state, it was enough to look at the rules which apply to
a quiescent cell and not to look at those which apply to a cell in a non-quiescent state 
although in the figures, in order to obtain nice pictures, we made implicit assumptions 
on rules applied to
a \hbox{{\tt B}-cell} or to a {\tt W}-cell behind the front whose neighbourhood may be
different from {\tt BW$^4$}, {\tt WBW$^3$} or {\tt B$^2$W$^3$}. If we ignore the
complex discussion involving a huge number of rules, we might expect an argument
on how long we have to wait for a new transformation of~$\nu$ from {\tt R} to {\tt B}.
Even if we have an argument on the number of possible configurations within
${\cal D}_{N_t}$, to repeat the same argument to $\sigma$ requires to consider
the number of possible configurations within ${\cal D}_{N_t+1}$ which is much bigger.
Accordingly, this leads to no conclusion, so that the case with three states is open,
even with rotation invariance.

Other questions may be considered. We know that strong universality is possible for 
deterministic cellular automata on the pentagrid with a quiescent state with ten states,
see~\cite{mmSU}. That cellular automaton is rotation invariant. What can be performed
if we relax rotation invariance? The answer is not straightforward as the cellular
automaton of~\cite{mmSU} is based on a cellular automaton on the line which is
strongly universal with eleven states and six states of that automaton could be
could be absorbed by the cellular automaton of the pentagrid which implements the
cellular automaton on the line. And so, for that direction two, a new approach is
needed.

Accordingly, as the gap between two states and ten states seems to be not a small one,
there is still a huge amount of work ahead.

\end{document}